\documentclass[10pt,conference,compsocconf,letterpape]{IEEEtran}
\usepackage{times, amsmath, amsfonts, graphicx, url, hyperref, xspace, shortcuts}
\usepackage{multirow}
\usepackage{amssymb}
\usepackage{pifont}
\usepackage{listings,multicol}
\usepackage{lipsum}
\usepackage{color}
\usepackage{semantic}
\usepackage[rounded]{syntax}
\usepackage{framed}

\usepackage{MnSymbol}

\usepackage{mathpartir}

\newcommand \bnfdef  {\mathrel{::=}}
\newcommand \bnfalt     {\mathrel{{\mid}}}
\newcommand{\xvdash}[1]{%
  \vdash_{\mkern-10mu\scriptscriptstyle\rule[-.9ex]{0pt}{0pt}#1}%
}

\definecolor{mygreen}{rgb}{0,0.6,0}
\definecolor{mygray}{rgb}{0.5,0.5,0.5}
\definecolor{mymauve}{rgb}{0.58,0,0.82}

\usepackage{ifpdf}
\ifpdf 
 \else \fi

\usepackage{rotating,enumerate}
\usepackage{algorithm}

\usepackage{algpseudocode}
\usepackage{subfig}
\usepackage{fancyvrb}
\DefineVerbatimEnvironment{smallcode}{Verbatim}{fontsize=\scriptsize,commandchars=\\\{\}}

\begin{document}

\ignore{
\author{
\IEEEauthorblockN{Xunchao Hu\IEEEauthorrefmark{1}, Yao Cheng\IEEEauthorrefmark{1}, Yue Duan\IEEEauthorrefmark{2}, Andrew Henderson\IEEEauthorrefmark{1}, Heng Yin\IEEEauthorrefmark{2}}

}
\IEEEauthorblockA{\IEEEauthorrefmark{1}Department of EECS, Syracuse University, USA 
\\\{xhu31, ycheng, anhender\}@syr.edu}
\IEEEauthorblockA{\IEEEauthorrefmark{2} Department of Computer Science and Engineering, University of California, Riverside
 \\\{yduan005, heng\}@cs.ucr.edu} 
}

\author{\IEEEauthorblockN{Xunchao Hu, Yao Cheng, Andrew Henderson}
	\IEEEauthorblockA{Department of EECS, \\Syracuse University, USA\\ 
	Email: \{xhu31, ycheng, anhender\}@syr.edu}
	\and
	\IEEEauthorblockN{Yue Duan, Heng Yin}
	\IEEEauthorblockA{ Department of CSE, \\University of California, Riverside \\
       Email: \{yduan005, heng\}@cs.ucr.edu}

}

%\title{\codename: Crash Free \& Framework Independent Execution of JavaScript}

%\title{JSBoost: Boosting Analysis of Malicious JavaScript via Forced Execution}
%\title{\codename: Uncover More Dangerous JavaScript via Forced Execution }
\title{JSForce: A Forced Execution Engine for Malicious JavaScript Detection}

\pagestyle{plain}
\pagenumbering{arabic}

\maketitle
\begin{abstract}

The drastic increase of JavaScript exploitation attacks has led to a strong interest in developing techniques to enable malicious JavaScript analysis. Existing analysis techniques fall into two general categories: static analysis and dynamic analysis. Static analysis tends to produce inaccurate results (both false positive and false negative) and is vulnerable to a wide series of obfuscation techniques. Thus, dynamic analysis is constantly gaining popularity for exposing the typical features of malicious JavaScript. However, existing dynamic analysis techniques possess limitations such as \ignore{the typical JavaScript malware is designed to execute in a particular environment since they aim to exploit specific vulnerabilities. This amplifies limitation of  the dynamic analysis  in terms of} limited code coverage and incomplete environment setup, leaving a broad attack surface for evading the detection. To overcome these limitations, we present the design and implementation of a novel JavaScript forced execution engine named \codename which drives an arbitrary JavaScript snippet to execute along different paths without any input or environment setup. We evaluate \codename using 220,587 HTML and 23,509 PDF real-world samples. Experimental results show that by adopting our forced execution engine, the malicious JavaScript detection rate can be substantially boosted by 206.29\% using same detection policy without any noticeable false positive increase. We also make \codename publicly available as an online service and will release the source code to the security community upon the acceptance for publication.

\hu{explain how 220,587 are calculated(malicius set plus benigh set)}
\end{abstract}

\section{Introduction}

Malicious JavaScript has become an important attack vector for software exploitation attacks. Attacks in browsers, as well as PDF files containing malicious embedded JavaScript, are typical examples of how attackers launch attacks using JavaScript. According to a recent report from Symantec~\cite{securityreport}, there are millions of victims attacked by malicious JavaScript on the Internet each day.

In recent years, a number of techniques~\cite{cova10:wepawet,Lu:2012,hartstein2009jsunpack,jsguard:gu, kolbitsch2012rozzle,nozzle:usenix09,curtsinger2011zozzle,kapravelos2013revolver, cao2014jshield} have been proposed to detect malicious JavaScript code. Due to the dynamic features of the JavaScript language, static analysis~\cite{feinstein2007caffeine, likarish2009obfuscated,seifert2008identification, curtsinger2011zozzle} can be  easily  evaded using obfuscation techniques~\cite{xu2012power}. Consequently, researchers rely upon dynamic analysis~\cite{cova10:wepawet,Lu:2012,hartstein2009jsunpack} to expose the typical features of malicious JavaScript. More specifically, these approaches rely upon visiting websites or opening PDF files with a full-fledged or emulated browser/PDF reader and then monitoring the different features ($eval$ strings~\cite{hartstein2009jsunpack}, heap health~\cite{nozzle:usenix09}, etc.) for detection.

 %Emulation based dynamic analysis techniques like Wepawet~\cite{cova10:wepawet}, jsunpack~\cite{hartstein2009jsunpack} are popular because they are easier to scale and provide far better control and visibility over the code execution. However, the limitation of code coverage and incorrect emulation of the targeted environment (Internet Explorer, Adobe PDF Reader, etc.) leave the attacker a broad attack surface~\cite{rajab2011trends} to bypass the detection.

However, the typical JavaScript malware is designed to execute within a particular environment, since they aim to exploit specific vulnerabilities, as opposed to benign JavaScript, which will run in a more environment-independent fashion. Fingerprinting techniques~\cite{upathilake2015classification} are widely adopted by JavaScript malware to examine the runtime environment. A dynamic analysis system may fail to observe some malicious behaviors if the runtime environment is not configured as expected. Such configuration is quite challenging because of the numerous possible runtime environment settings. Hence, existing dynamic analysis systems usually share the limitations of limited code coverage and incomplete runtime environment setup, which leave attackers with a broad attack surface to evade the analysis.

\ignore{The dynamic analysis system can not execute the sample with every possible environment setup since the combinations of browser/plugin/PDFReader is too many.  This results in the limited code coverage and incorrect runtime environment setup, which leaves attackers a broad attack surface to evade dynamic analysis.}

To solve those limitations, Rozzle~\cite{kolbitsch2012rozzle} explores multiple environment related paths within a single execution. But it requires a predefined environment-related profile for path exploration. Construction of a complete profile is a challenging task because of the numerous different browsers and plugins, especially for recent fingerprinting techniques~\cite{mowery2011fingerprinting, mulazzani2013fast}. These fingerprinting techniques do not rely upon any specific APIs, and thus Rozzle can be evaded because the predefined profile cannot include those fingerprinting techniques.  Also, Rozzle may introduce runtime errors because it executes infeasible paths  which may stop the analysis before the malicious code is executed. Revolver~\cite{kapravelos2013revolver} employs a machine learning-based detection algorithm to identify evasive JavaScript malware. However, it is dependent upon a known sample set and is unable to detect 0-day JavaScript malware. Although symbolic execution of JavaScript~\cite{saxena2010symbolic} can be applied to explore all of the possible execution paths, the performance overhead of a symbolic string solver~\cite{trinh2014s3} and the dynamic features of JavaScript make it infeasible for practical use.

\hu{Talk about the use case scenario, we don't rely on browser, or symbolic values}

In this paper, we propose \codename, a forced execution engine for JavaScript, which drives an arbitrary JavaScript snippet to execute along different paths without any input or environment setup. While increasing code coverage, \codename can tolerate invalid object accesses while introducing no runtime errors during execution.  This overcomes the limitations of current JavaScript dynamic analysis techniques. Note that, as an amplifier technique, \codename does not rely on any predefined profile information or full- fledged hosting programs like browsers or PDF viewers, and it can examine partial JavaScript snippets collected during an attack. As demonstrated in Section~\ref{sec:eval}, \codename can be leveraged to improve the detection rate of other dynamic analysis systems without modification of their detection policies. While the high-level concept of forced execution has been introduced in binary code analysis (X-Force~\cite{xforce}, iRiS~\cite{deng2015iris}), we face unique challenges in realizing this concept in JavaScript analysis, given that JavaScript and native code are very different languages by nature. 
%we solve several new challenges, such as type inference and path exploration, to enable forced execution of a JavaScript engine. 

We implement \codename on top of the V8 JavaScript engine~\cite{v8engine} and evaluate the correctness, effectiveness, and runtime performance of \codename with 220,587 HTML files and 23,509 PDF samples. Our experimental results demonstrate that adopting \codename can greatly improve the JavaScript analysis results by 206.29\% without any noticeable increase in false positives and with reasonable performance overhead.

Our main contributions are summarized as follows:

\begin{enumerate}
\item We propose forced execution of JavaScript, a technique that forces a JavaScript snippet to execute along different paths while requiring no inputs or any environment setup, to overcome the current limitations of existing JavaScript dynamic analysis techniques: limited code coverage and incomplete runtime environment setup.

\item To enable forced execution of JavaScript, we develop a type inference model to detect and properly recover from exceptions. We have also developed path exploration algorithms for malicious JavaScript code analysis. 

\item We implement the technique with a prototype system, named \codename, and evaluate its correctness, effectiveness, and runtime performance using 220,587 HTML and 23,509 PDF real-world samples. Experimental results show that by adopting \codename, the malicious JavaScript detection rate is substantially increased by 206.29\% while still using the same detection policy. This increase comes without any noticeable increase in false positives and with runtime performance that is very suitable for large-scale analysis.

\item We create an online service and make \codename publicly available at ~\cite{servicelink}. Upon the acceptance for publication, we will release the source code of \codename to the security community. 

\end{enumerate}

\section{Background and Overview}\label{bgoverview}
To provide the reader with a better understanding of the motivation for our system and the problems that it addresses, we begin with a discussion of the malicious JavaScript code used in drive-by-download attacks.

\begin{figure}
\centering
\begin{lstlisting}[frame=single]
if ((navigator.appName.indexOf("Microsoft Inte" + "rnet Explorer") == -1) && (navigator.userAgent.indexOf("Windows N" + "T 5.1") == -1) && (navigator.userAgent.indexOf("MSI" + "E 8.0") == -1)) {
    att = btt + 1;
}
if (att == 0) {
    try {
        new ActiveXObject("UM0QS4dD");
    } catch (e) {
        var tlMoOul8 = '\x25' + 'u9' + '\x30' + '\x39' + YYGRl6;
        tlMoOul8 += tlMoOul8;
        var CBmH8 = "%u";
        var vBYG0 = unescape;
        var EuhV2 = "BODY";
        ...    
    }
}
setTimeout("redir()", 3000);
\end{lstlisting}
\caption{The Malicious JavaScript Sample}
\label{fig:jssample}
\end{figure}

\hu{ modify the sample code to show the weakness of Rozzle(runtime error, predefined rozzle)}
\paragraph{Malicious JavaScript code}

\ignore{A typical m}Malicious JavaScript code is typically obfuscated and will attempt to fingerprint the version of the victim's software (browser, PDF reader, etc.), identify vulnerabilities within that software or the plugins that that software uses, and then launch one or more exploits. Figure~\ref{fig:jssample} shows a listing of JavaScript code used for a drive-by-download attack against \ignore{users of }the Internet Explorer browser. Line 1 employs precise fingerprinting to deliver only selected exploits that are most likely to successfully attack the\ignore{ client} browser. Lines 5-7 contain evasive code \ignore{which can }to bypass emulation-based detection systems. More precisely, the code attempts to load a non-existant ActiveX control, named {\tt UM0QS4dD} (line 6). When executed within a regular browser, this operation fails, triggering the execution of the {\tt catch} block that contains the exploitation code (lines 7-14). 

However, an emulation-based detection system \ignore{will}must emulate the ActiveX API by simulating the loading and presence of any ActiveX control. In these systems, the loading of the ActiveX control \ignore{does}will not raise this exception. As a result, the execution of the exploit \ignore{is stopped}never occurs and \ignore{and causes the system to fail to detect the}no malicious activity is observed. \ignore{Line 16 is used to redirect }Instead, the victim is redirected to a benign page (line 16) if the fingerprinting or evasion stage fails. Attackers can also abuse the function {\tt setTimeout} to create a time bomb~\cite{brumley2008automatically} to evade detection. \ignore{since the d}Detection systems can not afford \ignore{a long time of analysis for every}to wait for long periods of time during the analysis of each sample in an attempt to capture randomly triggered exploits.

\paragraph{Challenges and Existing Techniques}\label{sec:challenge}

\ignore{While s}Static analysis is a powerful technique that \ignore{allows one to }explores all \ignore{program }paths of execution. But, one particular issue that plagues static analysis \ignore{in the context }of malicious JavaScript is that not all of the code can be statically observed.  For \ignore{instance}example, static analysis cannot observe malicious code hidden within {\tt eval} strings, which are frequently exploited by attackers to obfuscate their code. Therefore, current detection approaches~\cite{cova10:wepawet,Lu:2012,hartstein2009jsunpack} rely upon dynamic analysis to expose features typically seen within malicious JavaScript.  More specifically, these approaches rely upon visiting websites or opening PDF files with an instrumented browser or PDF reader, and then monitoring different features ({\tt eval} strings~\cite{hartstein2009jsunpack}, heap health~\cite{nozzle:usenix09}, etc.) for detection.

However, dynamic analysis techniques suffer from two fundamental limitations. The first limitation is limited code coverage. \ignore{This is a well-known limitation of the dynamic analysis technique. It}This becomes a much more severe limitation within the context of analyzing malicious JavaScript. Attackers frequently employ a technique called $cloaking$~\cite{wang2011cloak}, \ignore{This technique}which works by fingerprinting the victim's web browser and only revealing the malicious content \ignore{only in case}when the victim is using a specific version of the browser with a vulnerable plugin. Cloaking makes dynamic analysis much harder because \ignore{defenders need to }the sample must be run within every combination of web browser and plugin to ensure complete code coverage. The widely-used event callback feature of JavaScript also makes it challenging for dynamic analysis to automatically trigger code. For example, attackers can load the attack code only when a specific mouse click event is captured, and automatically determining and generating such a trigger event is difficult.

The second limitation is the complexity of the JavaScript runtime environment.  JavaScript \ignore{language has been deployed }is used within many applications, and it can call the functionality of any plugin extensions supported by \ignore{many}these applications.\ignore{ like browsers.} For dynamic analysis, any pre-defined browser setup \ignore{can only }handles a known set of browsers and plugins. Thus, there is no guarantee that this setup will detect vulnerabilities only present in less popular plugins. While it is \ignore{feasible}possible to deploy a cluster of machines running many different operating systems, \ignore{as well as browser manufactures or versions or}browser applications, and browser plugins, the \ignore{combinational}exponential growth of possible \ignore{plugin version/browser/browser version}combinations rapidly \ignore{expands the requirements on hardware, network bandwidth and power.  This }causes scalability issues and \ignore{inefficient use of resources.}makes this approach infeasible.

Rozzle~\cite{kolbitsch2012rozzle} attempts to address this code coverage problem by exploring environment-related paths within a single execution. For instance,  because {\tt att} in Figure~\ref{fig:jssample} depends upon the environment-related API's output, Rozzle will execute lines 5-15 and reveal the malicious behaviors hidden in lines 8-14 by executing both the {\tt try } and {\tt catch} blocks. But, it requires a predefined environment-related profile for path exploration. Construction of a complete profile is a challenging task because of the numerous different browsers and plugins, especially for newer proposed fingerprinting techniques ~\cite{mowery2011fingerprinting, mulazzani2013fast, upathilake2015classification}. These new techniques do not rely upon any specific APIs. For instance, the JavaScript engine fingerprinting technique~\cite{mulazzani2013fast} relies upon JavaScript conformance tests such as the Sputnik~\cite{sputnik} test suite to determine a specific browser and major version number. There are no specific APIs used for the fingerprinting. Thus, Rozzle cannot include it within the predefined profile and explore the environment-related paths. Rozzle also introduces runtime errors into the analysis engine, which may stop the analysis before any malicious code is executed. In contrast, \codename does not rely upon predefined profile for path exploration and handles runtime errors using the forced execution model presented in Section~\ref{sec:forcedexecution}. By overcoming those limitations of Rozzle, \codename achieves greater code coverage. 

\ignore{
\ignore{To solve the code coverage problems, }Rozzle~\cite{kolbitsch2012rozzle} attempts to address this code coverage problem by exploring environment-related paths within a single execution. \ignore{But it 1) }But, it requires a predefined environment-related profile for path exploration, which may be incomplete. \ignore{and  2)}It also introduces runtime errors into the analysis engine, which may stop the analysis before any malicious code is executed. 
}

Revolver~\cite{kapravelos2013revolver} employs a machine learning-based detection algorithm to identify evasive JavaScript malware. However, it requires that the malicious sample is present within a \ignore{seen in }known sample set so that its evasive version can be determined based upon the classification difference. By design, it can not be used for 0-day malware detection.

\ignore{Researchers have tried applying techniques of s}Symbolic execution has also been applied to the task of exposing malware~\cite{brumley2008automatically}. This \ignore{approach}technique, while improving code coverage over dynamic analysis, suffers from scalability challenges and is, in many ways, unnecessarily precise~\cite{kolbitsch2012rozzle}. Within the context of JavaScript analysis,  symbolic execution becomes more challenging~\cite{saxena2010symbolic}. JavaScript applications accept many different kinds of input, and those inputs are structured as strings. For \ignore{instance}example, a typical application might take user input from form fields, messages from a server via {\tt XMLHttpRequest}, and data from code running concurrently within other browser windows. It is extremely difficult for a symbolic string solver~\cite{trinh2014s3} to effectively supply values for all of these different kinds of inputs and reason about how those inputs are parsed and validated. The rapidly evolving JavaScript language and its host programs (browsers, PDF readers, etc.) make the modeling of the JavaScript API tedious work. Furthermore, the dynamic features (such as the {\tt eval} function) of JavaScript make symbolic execution infeasible for many analysis efforts.

\paragraph{Overview} \ignore{In this paper, we propose }\codename, our proposed forced-execution engine for JavaScript, \ignore{ which, as its name suggests, forces an arbitrary JavaScript snippet to execute along different paths without any input or environment setup. \codename }is an enhancement technology designed to better expose the behaviors of malicious JavaScript at runtime. \ignore{so that the d}Different detection policies can be applied to examine \ignore{malice}malicious JavaScript. While the forced execution concept is first introduced for binary code analysis (X-Force~\cite{xforce}), we face unique challenges, such as type inference and invalid object access recovery, in enabling the forced execution concept for JavaScript. 

We now illustrate how the forced execution of JavaScript code works. Consider the snippet shown in Figure~\ref{fig:jssample}. \codename forces the execution through the different code paths of the snippet. So, the exploitation code within the {\tt catch} block (lines 7-14) will be executed, no matter how the ActiveX API is simulated by the emulation-based analysis system. Moreover, \codename will immediately invoke the callback function passed to {\tt setTimeout} to trigger the time bomb malware. 

\ignore{The path exploration of }\codename's path exploration forces line 2 to be executed, regardless of the result of the fingerprinting statement (line 1). Since {\tt btt} is not defined within the code snippet under analysis, which is a common scenario because collected JavaScript code may be incomplete due to multi-stages of the attack, the execution of line 2 raises a {\tt ReferenceError} exception when running within a normal JavaScript engine. When the exception is captured, \codename creates a {\tt FakedObject} named {\tt btt}, which is fed to the JavaScript engine to recover from the invalid object access. However, the type of {\tt btt} is unknown at the time of {\tt FakedObject}'s creation. \codename infers the type based upon how the {\tt FakedObject} is used. \ignore{So at runtime once }For example, if this {\tt FakedObject} is \ignore{used to }added to an integer, \codename will then change its type from {\tt FakedObject} to {\tt Integer}.  We call this \textit{faked object retyping}. 

%\textbf{Draw an architecture fig}

%While increasing the code coverage, \codename can tolerate the invalid object access and no runtime errors are introduced during the execution. This overcomes the limitation of current dynamic analysis techniques of JavaScript and can be leveraged to improve the detection rate without modification of the detection policy.  While the forced execution concept is not new in binary code analysis(X-Force~\cite{xforce}), we solve several new challenges like type inference, path exploration \textbf{details the challenges}  to enable forced execution on JavaScript engine.

\section{JavaScript Forced Execution}
\setcounter{paragraph}{0}

This section explains the basics of how a single forced execution proceeds. The goal is to have a non-crashable execution. \ignore{For readability, w}We first present the JavaScript language semantics and then focus on how to detect and recover from invalid object accesses. \ignore{ in this section. The other aspects of forced execution such as path exploration  are introduced in later sections.}We then discuss how path exploration occurs during forced execution.

\subsection{Forced Execution Semantics}\label{sec:forcedexecution}

\ignore{This section explains the basics of how a single forced execution proceeds. The goal is to have a non-crashable execution. For readability, we present the JavaScript language semantics first, and then focus on how to detect and recover from invalid object access in this section. The other aspects of forced execution such as path exploration  are introduced in later sections.}

\hu{add explaination of the type language in figure3, why is the object type needed?, make it clear this is dynamic typing and original V8's retyping cannot work in our scenario. what is type cast functions? }

\paragraph{The JavaScript Language}

\begin{figure}[ht]
\centering

\begin{small}
\begin{grammar}

<EXPRESSIONS> ::= $c$  \hfill CONSTANT
\alt $x$  \hfill VARIABLE
\alt $x.f$  \hfill FIELD ACCESS
\alt $x.prot$  \hfill PROTO ACCESS
\alt $e\ op\ e$  \hfill BINARY OP
\alt $this$  \hfill THIS
\alt \{$f_1:e_1,...,f_n:e_n$\}  \hfill OBJECT LITERAL
\alt \{function($p_1,...,p_n$)\{S\}\} \hfill FUNCTION DEF
\alt f($a_1,...,a_n$)	\hfill FUNCTION CALL
\alt new f($a_1,...,a_n$) \hfill NEW

<STATEMENTS> ::= skip \hfill SKIP
\alt $S_1:S_2$ \hfill SEQ
\alt $var x$ \hfill VAR DECL
\alt $x := e$ \hfill ASSIGN
\alt $x.f := e$ \hfill ASSIGN
\alt if $e$ then $S_1$ else $S_2$ \hfill CONDITIONAL
\alt while $e$ do $S$ \hfill WHILE
\alt try\{S\}catch\{S\}finally\{S\} \hfill TRY CATCH
\alt return $e$ \hfill RETURN

\end{grammar}

\end{small}

\caption{Core JavaScript}
\label{fig:jssyntax}
\end{figure}

JavaScript is a high-level, dynamic, untyped, and interpreted programming language. Figure~\ref{fig:jssyntax} summarizes the syntax of the core JavaScript, which captures the essence of JavaScript. At runtime, the JavaScript engine dynamically interprets JavaScript code to 1) load/allocate objects, 2) determine the types of objects, and 3) execute the corresponding semantics. %Generally, whenever there is a property access, the JavaScript engine traverses the heap for the pointer of the property. The failed traverse causes the reference error and abortion of the current execution. 
Given an arbitrary JavaScript snippet, execution may fail because of undefined/uninitialized objects or incorrect object types. For instance, the execution of line 2 in Figure~\ref{fig:jssample} raises a {\tt ReferenceError} exception because {\tt btt} is not defined. To tolerate such invalid object accesses, forced execution must handle such failures.\ignore{, which is explained in  the following paragraphs. }

\begin{figure}[ht]
\centering

\begin{small}

% \begin{multline*}
% \begin{align*}
\[
\begin{array}{rcl}
\text{Types:} \\
​
\tau &\bnfdef& \sum\nolimits_{i\in\emph{T,T}\subseteq\{\bot, u, b, s, n, o\}}\varphi_i  \\
\text{Rows:}  \\
\varrho &\bnfdef& \emph{str:}\tau,\varrho \\ 
		&\bnfalt& \varrho_\tau \\
\text{Type environments:} \\
\Gamma &\bnfdef& \Gamma(x:\tau) \\
	   &\bnfalt& {\emptyset}           \\
\text{Type summands} \\ 
\text{and indices:} \\
\varphi_\bot &\bnfdef& \text{Undef} \\
\varphi_u &\bnfdef& \text{Null} \\
\varphi_b &\bnfdef& \text{Bool}(\xi_b) \\
\xi_b &\bnfdef& false \bnfalt true \bnfalt \top \\
\varphi_s &\bnfdef& \text{String}(\xi_s) \\
\xi_s &\bnfdef& str \bnfalt \top \\
\varphi_n &\bnfdef& \text{Number}(\xi_n) \\
\xi_n &\bnfdef& num \bnfalt \top \\
\varphi_f &\bnfdef& \text{Function}(this: \tau;\varrho \to \tau) \\
\varphi_o &\bnfdef& \text{Obj}(\sum\nolimits_{i\in\emph{T,T}\subseteq\{b, s, n, f, \bot\}}\varphi_i)(\varrho) \\
\varphi_{fo} &\bnfdef & \text{FObj} \\
\varphi_{ff} &\bnfdef & \text{FFun}
% &\
% \end{align*}
\end{array}
\]
% \end{multline*} 

\end{small}

\caption{Syntax of JavaScript Types}
\label{fig:jstypes}
\end{figure}

%\note{discuss JavaScript object inheritance model}

The basic idea behind forced execution is that, whenever a reference error is discovered, a {\tt FakedObject} is created and returned as the pointer of the property. During the execution of the program, the expected type of the {\tt FakedObject} is indicated by the involved operation. For instance, adding a {\tt number} object to a {\tt FakedObject} indicates that the {\tt FakedObject}'s type is {\tt number}. When the type of a {\tt FakedObject} can be determined, we update it to the corresponding type. 

Potentially, we could assign {\tt FakedObject} with the type {\tt Object} and reuse the dynamic typing rules of the JavaScript engine to coerce the {\tt FakedObject} to an expected type. Nevertheless, the dynamic typing rules of the JavaScript engine are designed to maintain the correctness of JavaScript semantics and do not suffice to meet our analysis goal of achieving maximized execution. This can be attributed to two reasons. First, while the JavaScript engine can cast the {\tt FakedObject:Object} to proper primitive values, it cannot cast the {\tt FakedObject:Object} to proper object types. For instance, when a {\tt FakedObject} with the type {\tt Object} is used as a function object, the JavaScript engine will raise the {\tt TypeError} exception according to ECMA specification~\cite{ecmascript}. Second, the casting of {\tt FakedObject} to primitive values by the JavaScript engine can lead to unnecessary loss of precision. To understand why, consider the following loop:

\begin{lstlisting}
c = a/2;
for(i= c;i<10000;i++)
{ 
  memory[i] = nop + nop + shellcode;
}
\end{lstlisting}

Since {\tt a} is not defined, a {\tt FakedObject} will be created. With the built-in typing rule of the JavaScript engine, {\tt c} will be assigned the value {\tt NaN}. The loop condition {\tt $i<10000$ } will always evaluate to false.  Thus, the loop body, which contains the heap spray code, will never be executed. Although the path exploration of \codename will guarantee that the loop body will be executed once, without executing the loop 10,000 times, it will likely be missed by heap spray detection tools because of the small chunk of memory allocated on the heap. 

Therefore, to overcome the above two issues, \codename introduces two new types, {\tt FObj}  and {\tt FFun}, to the JavaScript type system. The JavaScript type system defined in~\cite{thiemann2005towards} is extended to support these two new types. Figure~\ref{fig:jstypes} summarizes the new syntax of these JavaScript types. Type {\tt FObj} is for {\tt FakedObject}. At the moment  {\tt FakedObject} is created, we assign type {\tt FObj} as the temporary type of {\tt FakedObject}. It can be subtyped to any types within the JavaScript type system.   When {\tt FakedObject} is used as a function object, {\tt FakedObject} is casted to {\tt FakedFunction} with type {\tt FFun}. The {\tt FakedFunction} with type {\tt FFun} can take arbitrary input and always returns {\tt FakedObject:FObj}. Following \codename's dynamic typing rules, {\tt a} in the above {\tt loop} sample will be typed to {\tt Number} because it is used as a dividend. {\tt c} is then assigned to {\tt Number} and the loop body is executed repeatedly until the loop condition {\tt i<10000} is evaluated to false. By introducing these two new types and their typing rules, \codename solves the two issues mentioned in the above paragraph. In the following paragraphs, we detail the JavaScript forced execution model. 

\begin{figure}
\centering
\begin{lstlisting}[multicols=2]
var a = null;
var b = c + 1;
var d = a.length;
var func = null;
a = "Hello World";
var e = new abc();
if (b < 5) {
    func = function(x) {
        return x
    };
}
d = func(6);
var f = Math.abs(d);
array[5] = f;
\end{lstlisting}
\caption{ JavaScript Sample}
\label{fig:jscase}
\end{figure}

\paragraph{Reference Error Recovery}To avoid raising {\tt ReferenceError} exceptions, we introduce the {\tt FakedObject} and recover the error by creating the {\tt FakedObject} whenever necessary. There are two cases that lead to reference errors. The first case (ER_1) is a failed object lookup. Every field access or prototype access triggers a dynamic lookup using the field or prototype's name as the key. If no object is found, the lookup fails. Such failures happen when the running environment is incomplete or some portion of the JavaScript code is missing. For example, a browser plugin \ignore{ like Adobe Acrobat are} referenced by the JavaScript is not installed, or only a portion of the JavaScript code is captured during the attack. 

To handle this error,  \codename intercepts the lookup process and a {\tt FakedObject} named as the lookup key is created whenever a failed lookup is captured. The corresponding parent object's property is also updated to the {\tt FakedObject}. Line 2 in Figure ~\ref{fig:jscase} presents such an example. The JavaScript engine searches the current code scope for the definition of {\tt c}, which is not defined. \codename returns the {\tt FakedObject} as the temporary value of {\tt c} so that the execution can continue.

The second case (ER_2) occurs when the object is initialized to the value {\tt null} or {\tt undefined}, but later has its properties accessed. \codename modifies the initialization process to replace the {\tt null} to a {\tt FakedObject} if an object is initialized as value {\tt null} or {\tt undefined}. For example, the variable {\tt a} defined on line 1 in Figure~\ref{fig:jscase} is assigned the value {\tt FakedObject} instead of {\tt null} under the forced execution engine.  The variable {\tt a} may later be updated to another value during execution, but this does not sabotage the execution of JavaScript code. 

\ignore{
\begin{figure}
\centering

\begin{grammar}

<PostfixOperator>	::=	"++" | "--"

<UnaryOperator>	::=	 "delete" | "void" | "typeof" | "++" | "--" | "+" | "-" | "~" | "!" 

<BitwiseOperator> := "&" | "^" | "|"

<MultiplicativeOperator>	::=	 "*" | "/"  | "%"

<AdditiveOperator>	::=	"+" | "-" 

<ShiftOperator>	::=	 "<<" | ">>" | ">>>" 

<RelationalOperator> ::= "<" | ">" | "<=" | ">=" | "instanceof" | "in" 

<EqualityOperator>	::=	 "==" | "!=" | "===" | "!==" 

<LogicalOperator>	::=	"&&" | "||"

<AssignmentOperator>	::=	 "=" | "*=" | "/=" | "%=" | "+=" | "-=" | "<<=" | ">>=" | ">>>=" | "&=" | "^=" | "|=" 

\end{grammar}

\caption{The JavaScript Operators}
\label{fig:JSOperators}
\end{figure}
}

\begin{figure*}

\small
\centering

\begin{mathpar}
\inferrule[R-ASSIGN]{\Gamma \xvdash{lhs} e_0: \varphi_{fo} \\ \Gamma \vdash e_1:\tau}{\Gamma \xvdash{ref} e_0 = e_1:\tau} \ \ \ \    
\inferrule[R-CALL1]{\tau_0 \unrhd Obj(Function(this:\tau^\prime;\lceil0 \rceil:\tau_1,...,\lceil n\textendash1\rceil:\tau_n,\varrho \rightarrow \tau  ))(\varrho^\prime) \\\\\Gamma \xvdash{ref} e_0: \varphi_{fo}/\tau }{ \xvdash{upd} e_0:\varphi_{fo}, \varrho^\prime @ \tau \leftmapsto \varphi_{ff} ,\Gamma \xvdash{ref} e_0(e_1,....,e_n):\varphi_{fo}/\perp}\\
\inferrule[R-CALL2]{\tau_0 \unrhd Obj(Function(this:\tau^\prime;\lceil0 \rceil:\tau_1,...,\lceil n\textendash1\rceil:\tau_n,\varrho \rightarrow \tau  ))(\varrho^\prime) \\\\\Gamma \xvdash{ref} e_0: \tau_0/\tau^\prime  \\ \Gamma \vdash e_1:\tau_1 \\...\Gamma \vdash e_{(i\textendash1)}:\tau_{(i\textendash1)}\\\Gamma \vdash e_i:\varphi_{fo} \\ \Gamma \vdash e_{(i+1)}:\tau_{(i+1)} \\...\\ \Gamma \vdash e_n:\tau_n }{ \xvdash{upd} e_i:\varphi_{fo} @ \tau \leftmapsto \tau_i ,\Gamma \xvdash{ref} e_0(e_1,....,e_n):\tau/\perp}\\
\inferrule[R-NEW]{\tau_0 \unrhd Obj(Function(this:\tau^\prime;\lceil0 \rceil:\tau_1,...,\lceil n\textendash1\rceil:\tau_n,\varrho \rightarrow \tau  ))(\varrho^\prime) \\\\\Gamma \xvdash{ref} e_0: \varphi_{fo}/\tau  \\  }{ \xvdash{upd} e_0:\varphi_{fo}, \varrho^\prime @ \tau \leftmapsto \varphi_{ff} ,\Gamma \xvdash{ref} new\  e_0(e_1,....,e_n):\varphi_{fo}/\perp}\ \ \ \ \ 
\inferrule[R-BINOPERATOR1]{\Gamma \vdash e_1:\varphi_{fo} \\ \Gamma \vdash e_2:\tau^\prime \\ \neg(e_2 \ is\ \varphi_{fo})  }{ \xvdash{upd}e_1:\varphi_{fo}@ \tau \leftmapsto \tau^\prime, \Gamma \vdash e_1\ op\ e_2:\tau^\prime}\\
\inferrule[R-BINOPERATOR2]{\Gamma \vdash e_1:\varphi_{fo} \\ \Gamma \vdash e_2:\varphi_{fo} }{ \xvdash{upd}e_1:\varphi_{fo}@ \tau \leftmapsto \varphi_{n},  \xvdash{upd}e_2:\varphi_{fo}@ \tau \leftmapsto \varphi_{n}, \Gamma \vdash e_1\ op\ e_2:\tau}\ \ \ \ \ 
\inferrule[R-INDEX1]{ \Gamma \vdash e_1 : \varphi_{fo} \\ \tau_1 \unrhd Obj(\varphi_1)(\varrho_1) \\ \Gamma \vdash e_2:\varphi_{n} \\ }{ \xvdash{upd} e_1:\varphi_{fo}@ \tau \leftmapsto \tau_1, \Gamma \xvdash{lhs} e_1[e_2]:\varphi_{fo}} \\
\inferrule[R_UNARYOPERATOR]{\Gamma \vdash e_1:\varphi_{fo} }{\xvdash{upd}e_2:\varphi_{fo}@ \tau \leftmapsto \varphi_{n}, \Gamma \vdash op\ e_1:\tau}\ \ \ 
\inferrule[R-INDEX2]{ \Gamma \vdash e_1 : \tau_1 \\ \tau_1 \unrhd Obj(\varphi_1)(\varrho_1) \\ \Gamma \vdash e_2:\varphi_{fo} \\ \xvdash{upd} \varrho_1@\varphi_{n} \mapsto \tau^\prime}{ \xvdash{upd} e_2:\varphi_{fo}@ \tau \leftmapsto \varphi_{n}, \Gamma \xvdash{lhs} e_1[e_2]:\tau^\prime}
\end{mathpar}

\caption{Typing rules\\}
\label{fig:typingrule}
\end{figure*}

\paragraph{Faked Object Retyping} 
When a {\tt FakedObject} is used within an expression, it must be retyped to the expected type. Otherwise, incorrect typing raises a {\tt TypeError} exception and stops the execution. \codename infers the expected type of {\tt FakedObject} by how the {\tt FakedObject} is used.  Figure~\ref{fig:typingrule} summarizes the dynamic typing rules introduced by \codename. The rules are divided into the following five categories:

\begin{enumerate}

\item {\em R-ASSIGN}. This rule deals with assignment statements. When a {\tt FakedObject} $e_0$  is assigned to a new value $e_1$, $e_0$ is updated to the new value $e1$ with the type $\tau$. The JavaScript engine handles this naturally, so no interference is required. For example, variable {\tt a} in Figure~\ref{fig:jscase} is assigned {\tt FakedObject} at line 1 by \codename. At line 4, the variable {\tt a} is retyped as a {\tt string} object.

\item {\em R-CALL1} and {\em R-NEW}.  These two rules describe the typing rule for the scenario when a {\tt FakedObject:FObj} is used as a function call or by the {\tt new} expression. Function calls and the {\tt new} expression both expect their first operand to evaluate to a function. So, \codename updates the {\tt FakedObject:FObj} to {\tt FakedFunction:FFun} for this situation. The {\tt FakedFunction} is a special function object  which is configured to accept arbitrary parameters. The return value of the function is set to a {\tt FakedObject:FObj} so that it can be retyped whenever necessary. 

\item {\em R-CALL2}. This rule describes the case where the callee is a known function, but a {\tt FakedObject:FObj} is passed as a function parameter. \codename types the {\tt FakedObject:FObj} to the required type of the callee's arguments. The JavaScript language has many standard built-in libraries such as {\tt Math} and {\tt Date}. When a {\tt FakedObject:FObj} is used by the standard library function, we update the type based upon the specification of the library function~\cite{ecmascript}. Currently,  \codename implements retyping for several common libraries (e.g., {\tt Math}, {\tt Number}, {\tt Date}). 

\item {\em R-BINOPERATOR1/2} and {\em R-UNARYOPERATOR}. These three rules describe how to update the type if the {\tt FakedObject:FObj} is involved in an expression with an operator. \codename updates the {\tt FakedObject:FObj}'s type based upon the semantics of the operator.  For unary operators, it is straightforward to determine the type from the operator's semantics. For instance, the postfix operator indicates the type as {\tt number}. For binary operators, the typing becomes more complicated.  If both operands are {\tt FakedObject:FObj} and the operator does not reveal the type of the operands, \codename types them to {\tt number}. This is because the {\tt number} type can be converted to most types naturally by the JavaScript engine.  For example, the {\tt number} type in JavaScript can be converted to the {\tt string} type, but it may fail to convert a {\tt string} to a {\tt number}. Later during execution, if the types can be determined, \codename will update the type to the correct type. If only one of the two operands is {\tt FakedObject:FObj}, \codename determines the type based upon the other operand's type and the operator's semantics.
 
\item {\em R-INDEX1} and {\em R-INDEX2}. These two rules describe how to update the type when there are indexing operations.  A {\tt FakedObject:FObj} is updated to an $Array Object:\phi_o$ whenever a key is used as an array index to access elements of the {\tt FakedObject}.  \codename creates an $Array Object$ and initializes the elements to {\tt FakedObject:FObj}. The length of the $Array Object$ is set to  $2$*$CurrentIndex$. If an Out-Of-Boundary access is found, \codename doubles the length of  $Array Object$. If the array index is {\tt FakedObject}, \codename types it to {\tt number} and initializes it as {\tt 0}, which avoids Out-Of-Boundary exceptions. If both the index object and base object are {\tt FakedObject:FObj}, the R-INDEX2 rule is first applied to update the index object to {\tt number}, then the R-INDEX1 rule is applied to update the base object to $ArrayObject$.

\end{enumerate}

\ignore{
\begin{table}[t]

\small
\centering
\begin{tabular}{|l|l|l|l|}
\hline

{\bf Line\#}&{\bf Statement} & {\bf Action} & {\bf rule}  \\
\hline\hline
1& var a = null; &  &  \\ \hline
2& var b = c + 1; &  &  \\ \hline
3& var d = a.length; & & \\ \hline
4& var func = null; & & \\ \hline
5&  a = "Hello World";& & \\ \hline
6&  var e = new abc();& & \\ \hline
7&  if (b $<$ 5) ;& & \\ \hline
12&  d = func(6); & &\\ \hline
13& var f=Math.abs(d)& & \\ \hline
14&  array[5] = f;& & \\ \hline
\end{tabular}
\caption{Forced execution  of sample in Figure~\ref{fig:jscase}}
\label{table:fesample}
\end{table}
}

\begin{table}[t]

\centering

\begin{tabular}{|l|l|l|l|}
\hline

Statement                           & Action                   & \multicolumn{2}{l|}{Rule} \\ \hline
 {\tiny\color{mygray}{1:}} var a = null;                       & $a \leftmapsto FakedObject$           & \multicolumn{2}{l|}{ER_2}     \\ \hline
\multirow{2}{*}{{\tiny\color{mygray}{2:}} var b = c + 1;}     & $c \leftmapsto FakedObject$  & \multicolumn{2}{l|}{ER_1}     \\ \cline{2-4} 
                                                         & $c \leftmapsto RanNumber$               & \multicolumn{2}{l|}{\specialcell{R_BINOPE\\RATOR1}}     \\ \hline
{\tiny\color{mygray}{3:}} var d = a.length;                   & $a.length \leftmapsto FakedObject$   & \multicolumn{2}{l|}{ER_1}     \\ \hline
 {\tiny\color{mygray}{4:}} var func = null;                    & $func \leftmapsto FakedObject$       & \multicolumn{2}{l|}{ER_2}     \\ \hline
 {\tiny\color{mygray}{5:}} a = "Hello World";                  & $a \leftmapsto "Hello World"$                & \multicolumn{2}{l|}{R_ASSIGN}     \\ \hline
 \multirow{2}{*}{{\tiny\color{mygray}{6:}} var e = new abc();} & $abc \leftmapsto FakedObject$        & \multicolumn{2}{l|}{ER_1}     \\ \cline{2-4} 
                                     & $abc \leftmapsto fakedFunction$ & \multicolumn{2}{l|}{R_NEW}     \\ \hline
 {\tiny\color{mygray}{7:}} if(b \textless5)                    & NO ACTION                & \multicolumn{2}{l|}{NONE}     \\ \hline
 \multirow{2}{*}{{\tiny\color{mygray}{12:}} d = func(6)}        & $func \leftmapsto fakedFunction$ & \multicolumn{2}{l|}{R_CALL1}     \\ \cline{2-4} 
                                                   & $d \leftmapsto FakedObject$          & \multicolumn{2}{l|}{R_ASSIGN}     \\ \hline
 {\tiny\color{mygray}{13:}} var f = Math.abs(d)                 & $d\leftmapsto RanNumber$       & \multicolumn{2}{l|}{R_CALL2}     \\ \hline
\multirow{3}{*}{{\tiny\color{mygray}{14:}} array{[}5{]} = f;}  & $array \leftmapsto FakedObject$    & \multicolumn{2}{l|}{ER_1}     \\ \cline{2-4} 
                                     & $array \leftmapsto arrayObject$         & \multicolumn{2}{l|}{R_INDEX1}     \\ \cline{2-4} 
                                     & $array{[}5{]} \leftmapsto f$  & \multicolumn{2}{l|}{R_ASSIGN} \\ \hline
\end{tabular}

\caption{Forced execution  of sample in Figure~\ref{fig:jscase}}
\label{table:fesample}

\end{table}

\paragraph{Example} Table~\ref{table:fesample} presents a forced execution of the sample shown in Figure~\ref{fig:jscase}. In the execution, the branch in lines 8-11 is not taken. At line 1, \codename assigns a {\tt FakedObject:Fobj} to {\tt a}, instead of {\tt null}. This is because at line 3 the access to property {\tt length} raises an exception if {\tt a} is {\tt null}.  On line 2, we can see a {\tt FakedObject:FObj} is first assigned to {\tt c}. Once {\tt c} is added to {\tt 1}, \codename updates the value of {\tt c} to a random number. Lines 6 and 7 show that if a {\tt FakedObject:FObj} is used in the function call or {\tt new} expression, \codename updates it to {\tt FakedFunction:FFun}. The return value of the faked function is still configured to {\tt FakedObject:FObj}, so that at line 13, {\tt d} is updated to hold a random number.

\ignore{

\begin{figure}
\centering
\begin{grammar}
<STATEMENT> := while $e$ do $S$ \hfill WHILE
\alt if $e$ then $S_1$ else $S_2$ \hfill CONDITIONAL
\alt try\{S\}catch\{S\}finally\{S\} \hfill TRY CATCH
\alt switch(e)\{case e: S ... default:S \} \hfill SWITCH
\alt e\ ?\ a\ :\ b \hfill TERNARY EXPRESION

\end{grammar}
\caption{The JavaScript Branch Statement}
\label{fig:branchstatement}
\end{figure}

}

\codename also automatically recovers from other exceptions by intercepting those exceptions to eliminate the exception condition.  For example, \codename will update a divisor to a non-zero value if a division-by-zero exception is raised.\ignore{  Details are omitted.}

\renewcommand{\algorithmicrequire}{\textbf{Input:}}
\renewcommand{\algorithmicensure}{\textbf{Output:}}
\begin{algorithm}[h!]

\caption{Path Exploration Algorithm}
\label{algorithm:pathexp}
{\small
\textbf{Definitions:} $switches$ - the set of switched predicates in a forced execution, denoted by a sequence of  predicate offsets in the source file(SrcName:offset). For example, $t.js:15 \cdot t.js:83 \cdot t.js:100$ means the branch in  source file $t.js$ with the offset 15, 83, 100  is switched.  $EX$,  $WL$ - a set of forced executions, each denoted by a sequence of switched predicates. $preds:\overline{Predicate \times boolean}$ - the sequence of executed predicates.

\begin{algorithmic}[1]
	\Require The tested $JS$
	\Ensure  $FULL\_EX $
	\State $FULL\_EX \gets \emptyset$
	\State $SRC \gets \{JS\}$
	\While {$SRC$}
	\State $WL \gets \{\emptyset\}$
	\State $EX \gets \emptyset$
	\State $js \gets SRC.pop()$
	\While {$WL$}
	\State $switches \gets WL.pop()$
	\State $EX \gets EX \cup switches$
	
	\State $(preds, newJS) \gets $ \Call{ExecuteCode}{$js,switches$}
	\State $SRC \gets SRC \cup newJS$
	\State $t \gets len(switches)$
	\State $preds \gets remove\ the\ first\ t\ elements\ in\ preds$

	\ForAll {$(p,b) \in preds $}
	\If {$!covered(p,\neg b)$}
	\State $WL \gets WL \cup switches \cdot (p,b)$
	\EndIf
	
	\EndFor
	\EndWhile
	\State $FULL\_EX \gets FULL\_EX \cup \{EX:js\} $
	\EndWhile

	\Procedure {ExecuteCode}{$JS, switches$}
	\State $preds\gets switches$
	\State $CBQ \gets \emptyset$
	\State $newJS \gets \emptyset$
	\ForAll {$stmt \in JS$}
		\If {$isNoneEvalFunctionCallStmt(stmt)$}
			\If {$CalleeTakesStrings(stmt)$}
			\State $newJS \gets newJS \cup GetJSFromString(stmt)$
			\EndIf
			\If {$CalleeRegisterCallback(stmt)$}
			\State $CBQ \gets CBQ \cup ExtractCBFunc(stmt)$
			\EndIf
		\ElsIf {$isBranchStmt(stmt)$}
			\If {$GetSwitch(stmt) \in switches$}
			\State $Execute\ according\ to\ switches$
			\Else
			\State $ preds \gets preds \cdot GetPredicate(stmt)$
			\EndIf

		\EndIf

	\EndFor

	\ForAll {$cb \in CBQ$}
		\State $(preds^\prime, newJS^\prime) \gets$ \Call {ExecuteCode}{$cb, \emptyset$}
		\State $newJS \gets newJS \cup newJS^\prime$
		\State $preds \gets preds \cdot preds^\prime$
	\EndFor

	\Return $(preds, newJS)$
	\EndProcedure

\end{algorithmic}

}

\end{algorithm}

\subsection{Path Exploration in \codename}

One important functionality of \codename is the capability of exploring different execution paths of a given JavaScript snippet to expose its behavior and acquire complete analysis results. In this subsection, we explain the path exploration algorithm and strategies.

In practice, attackers constantly adopt the dynamic features of JavaScript to aid in evading detection. This results in incomplete path exploration under two circumstances. The first is when strings are dynamically generated. For instance, {\tt document.write} is often abused to inject dynamically decoded malicious JavaScript code into the page at runtime. The second is when event callbacks are used. As discussed in Section~\ref{bgoverview}, attackers can abuse event callbacks to stop the execution of malicious code. \codename solves this by employing specific path exploration strategies. Within the execution, if faked functions take strings as input, \codename examines the strings and executes the code  if they contain JavaScript. This strategy is only applied on faked functions since original functions ({\tt eval}) can handle the strings as defined. \codename also detects the callback registration function and invokes the callback function immediately after the current execution terminates.

%The forced execution on x86 code only needs to consider conditional jump instruction to flip the predicates. 

\ignore{JavaScript has several branch statements.  Particularly, }\codename treats {\tt try-catch} statements as {\tt if-else} statements, ie., it executes each {\tt try} block and {\tt catch} block separately. Ternary operators are also treated as {\tt if-else} statements: both values are evaluated.

There are several different path exploration algorithms: linear search, quadratic search, and exponential search~\cite{xforce}. The goal of path exploration in \codename is to maximize the code coverage to improve the detection rate of malicious payload with \ignore{decent} an acceptable performance overhead. Quadratic and exponential searches \ignore{ cost too much}are too expensive, so \codename employs the linear search only. 

Algorithm~\ref{algorithm:pathexp} describes the path exploration algorithm, which generates a pool of forced executions that achieve maximized code coverage. The complexity is $O(n)$, where $n$ is the number of JavaScript statements. $n$ may change at runtime because JavaScript code can be dynamically generated. Initially, \codename executes the program without switching any predicates since {\tt switches} is initialized  as $\emptyset$ (line 8) for the first time. \codename executes the program according to the {\tt switches} at line 10 and returns {\tt preds} and dynamically generated code {\tt newJS}. In lines 12-17, we determine if it would be of interest to further switch more predicate instances. Lines 11-13 compute the sequence of predicate instances eligible for switching. Note that it cannot be a predicate before the last switched predicate specified in {\tt switches}. Switching such a predicate may change the control flow such that the specification in {\tt switches} becomes invalid. Specifically, line 16 switches the predicate if the other branch has not been covered. In each new forced execution, we essentially switch one more predicate.

The  procedure {\tt ExecuteCode} (lines 22-47) describes the execution process. It collects dynamically generated JavaScript code (lines 28-30) and the executed predicates (lines 34-38). The new generated JavaScript code, {\tt newJS}, will be executed after the path exploration of the current {\tt js} finishes. The registered callback functions (lines 31-33) are also queued and invoked after the current execution finishes (lines 42-46). As an example, recall the callback function {\tt redir()} used in line 16 of Figure~\ref{fig:jssample}. Instead of waiting for the timeout, \codename will trigger the {\tt redir()} function immediately after the current execution finishes.

\section{Implementation}
\setcounter{paragraph}{0}

\codename  is implemented by extending the V8 JavaScript engine~\cite{v8engine} on the X86-64 platform. It is comprised of approximately 4,600 lines of C/C++ code and 1,500 lines of Python code. We address some prominent challenges of its implementation in this section.

\paragraph{Reference Error Recovery \& Faked Object Retyping}
In V8, an abstract syntax tree (AST) is generated for every function, which is then compiled into native code (known as Just-In-Time code). V8 adopts an inline caching technique~\cite{holzle1991optimizing} to accelerate property accesses. If the property access fails, the execution jumps to the V8 runtime system which handles any inline cache misses. If the runtime system is unable to handle an inline cache miss, either due to reference error or type check error, it raises the corresponding exception and stops the execution.
	
We modify the inline cache miss handling process to enable reference error recovery and faked object retyping. For reference error recovery, \codename creates and returns the {\tt FakedObject} for failed object lookup by changing the V8 property access failure handling functions like {\tt Runtime_LoadIC_Miss}. For faked object  retyping, \codename inserts additional code into runtime methods like {\tt Runtime_BinaryOpIC_Miss} that is executed prior to the exception being raised. This additional code follows the rules described in Section~\ref{sec:forcedexecution} to conduct the retyping process if the involved operation contains a {\tt FakedObject}. 

\paragraph{Predicates Flip}

We have two approaches available to flip the predicates. The first approach is to flip the predicates within the Just-in-Time code. The Just-in-Time code can be optimized (inline caching, etc.) by V8 in accordance with the execution profile. To enable predicates flipping, a runtime function must be inserted before every branch so that \codename can manipulate the predicate value. This approach may affect the optimization process of Just-in-Time code. 

%\andrew{Paragraph on second option, AST modification, goes here.}

\codename takes the second approach: if the branch $A$ of a predicate needs  to be taken, \codename replaces the other branch with this branch $A$. At runtime, no matter which branch is taken, the branch $A$ is executed.  For instance, we want to take the $\{A\}$ branch of the statement {\tt if(e)\{}$A${\tt \}else\{}$B${\tt \}}. \codename  changes it to {\tt if(e)\{}$A${\tt \}else\{}$A${\tt \}}, so that $\{A\}$ is executed at runtime.

\paragraph{Loops and Recursions} %\codename may corrupt variables, and if a loop condition variable is corrupted, an (incorrect) infinite loop may result. 
Sometimes, \codename may cause a loop to execute for a very long time, due to the introduction of faked objects. To solve this problem,  \codename inserts a time counter for every loop statement ({\tt for...in} and {\tt for...of} are excluded, as they will always terminate), and it will terminate the loop if the execution time exceeds a limit. Similarly, if \codename forces a predicate that guards the termination of a recursive function call, a very deep recursion may result. To address deep recursion, \codename monitors the stack depth. Once the maximum call stack size (defined by V8) is reached, calls to that function are omitted by \codename.

\section{Evaluation}\label{sec:eval}

\setcounter{paragraph}{0}

In this section, we present details on the evaluation of correctness, effectiveness and runtime performance of \codename using a large number of real-world samples.

%\codename is implemented on top of the V8 JavaScript engine~\cite{v8engine}. It comprises XX lines of C/C++ code and XX lines of Python code. We evaluate the cost and benefit of \codename by comparing original jsunpack~\cite{hartstein2009jsunpack} against enhanced jsunpack with \codename.

\subsection{Dataset \& Experiment Setup}

\paragraph{Dataset}
The complete dataset used for our evaluation consists of two sample sets: a malicious sample set and a benign sample set. For the malicious set, we collected a sample set with 172,995 HTML files and 23,509 PDF files from various databases including VirusTotal~\cite{virustotal}, Contagio~\cite{contagio}, MalTrafficAnalysis~\cite{maltraffic}, and Threatglass~\cite{threatglass}. Among those, all samples from VirusTotal were new samples evaluated within a month of being submitted, with the samples provided from other sources being relatively old. For the benign sample set, we crawled the Alexa top 100 websites~\cite{alexa} and collected 47,592 HTML files.

\paragraph{Experiment Setup}
For JavaScript code analysis, we leverage the jsunpack~\cite{hartstein2009jsunpack} tool. Jsunpack is a widely used malicious JavaScript code analysis tool that utilizes the SpiderMonkey~\cite{spidermonkey} JavaScript engine for code execution. Six distinct configurations are predefined within jsunpack to maximize the exploration of JavaScript code by trying different browsers and language settings. For the sake of our evaluation, we replaced the SpiderMonkey from jsunpack with \codename and relied upon the detection policies in jsunpack for malicious code detection. Most of our experiments are based upon the comparison between the original jsunpack and the \codename-extended jsunpack. Note that the experiments performed within this paper are only intended to show the \textit{improvement of detection results} over the original ones when adopting \codename. The detection policy itself is another important research topic which is orthogonal to the focus of this paper. We conducted our experiments on a test machine equipped with Intel(R) Xeon(R) E5-2650 CPU (20M Cache, 2GHz) and 128GB of physical memory. The operating system was Ubuntu 12.04.3 (64bit).

%\subsection{Improved Detection Rates with \codename}

%\paragraph{Setup:} We ported the detection rule deployed in jsunpack~\cite{hartstein2009jsunpack} to \codename. Then we ran both on the data set to compare the detection rates.

\subsection{Correctness}

\setcounter{paragraph}{0}

In this section, we evaluate the correctness of the analysis result for \codename. The goals of this evaluation are two-fold. First, we wish to know the true positive rate of our analysis results, meaning that we wish to verify whether a JavaScript program is undoubtedly malicious if it is tagged as one by the analysis tools. Second, we wish to understand any false positives in the results so as to determine whether any benign JavaScript code can be mistakenly labeled as malicious.

%\hu{Remove comparison with jsunpack in table 2 correctness results. and show the 600~samples analysis results.}

\begin{table}[t]
%\vspace{-0.1in}
\small
\centering
\resizebox{\linewidth}{!}{
\begin{tabular}{|l|l|l|l|}
\hline
{\bf Category} & {\bf Total} & {\bf Detected by \codename} & {\bf Percentage} \\
\hline\hline
True Positive & 389 & 389 & 100\%\\
\hline%\hline
False Positive & 47,592 & 9 & 0.019\%\\
\hline
\end{tabular}}
\caption{Correctness Results.}
\label{table:correctness}
\vspace{-0.1in}
\end{table}

\paragraph{True positive} With our first goal in mind, we queried VirusTotal~\cite{virustotal} for malicious HTML files and collected 389 samples which are precisely labeled with specific CVE (Common Vulnerabilities and Exposures) numbers that match CVEs listed in jsunpack. Furthermore, we manually reviewed each of the samples and confirmed the existence of shellcode or malicious signatures. This step is to guarantee all the samples we tested are real malicious samples that should be detected by our tool. Then, we analyzed the samples using jsunpack with \codename. The experimental result is listed in the first row of Table~\ref{table:correctness} as ``true positive''. It shows that \codename could successfully detect all of the samples, resulting in a 100\% true positive rate. To better understand these results, we further inspected the detailed analysis results to see why our tool tagged samples as malicious. Our inspection results revealed that all of the payload and malicious signatures extracted by the \codename are indeed malicious, proving that our tool can achieve very high true positive rate with accurate analysis details. %As for the undetected samples, the \codename-extended tool is able to explore all of the branches within the code. The reason why some samples can not be detected is that some of the sample programs contained syntax errors that halted the JavaScript's execution.

%With our first goal in mind, we manually selected 190 malicious samples which are precisely labeled with specific CVE (Common Vulnerabilities and Exposures) numbers that match CVEs listed in jsunpack. Furthermore, we reviewed each of the samples and confirmed the existence of shellcode or malicious signatures. This step is to guarantee all the samples we tested are real malicious samples that should be detected. Then, we analyzed the samples using jsunpack with \codename. The experimental results are listed in the first row of Table~\ref{table:correctness} as ``true positive''. It shows that \codename could detect a total of 180 malicious samples out of 190 . To better understand these results, we further inspected every detected and undetected sample. Our inspection results revealed that all of the payload detected by the \codename-extended version are truly payload codes. As for the undetected samples, the \codename-extended tool is able to explore all of the branches within the code. The reason why some samples can not be detected is that some of the sample programs contained syntax errors that halted the JavaScript's execution.

\paragraph{False positive} For our second goal, we analyzed our benign sample set using \codename and then observed whether any of the samples could be incorrectly labeled as malicious. As shown in the second row of Table~\ref{table:correctness}, the \codename tags 9 out of 47,592 samples as malicious. We first manually confirmed that all 9 samples are clean and thereupon study why the false positives happen. It has been verified by manual inspection that all of the false positives are caused by the inaccurate detection policy, to be more specific, the over-relaxation of the shellcode string matching policy enforced by jsunpack. The reason why our tool could detect them as malicious is that it explores JavaScript code in a more complete fashion in consequence of our forced execution technique. Therefore, based upon the above experimental results, we argue that using \codename will keep a very low false positive rate for JavaScript code analysis, and is able to assist in accomplishing more thorough results. Theoretically, \codename can generate higher code coverage than jsunpack and lead to better analysis results. But, the question is by how much. With that, we conducted another set of experiments to show the effectiveness of \codename.

\ignore{
\begin{table}[t]
%\vspace{-0.1in}

\centering
%\resizebox{\linewidth}{!}{
\begin{tabular}{|l|l|l|l|l|}
\hline
{\bf Sample Set} & {\bf Total} & \specialcell{\bf without \\ \codename} & \specialcell{\bf with \\ \codename} &{\bf Improvement}\\
\hline\hline
Old HTML & 66,325 & 193 & 357 & 84.9\%\\
\hline%\hline
New HTML & 106,018 & 2,250 & 20,649 & 817.3\%\\
\hline%\hline
{\bf HTML Total} & {\bf 172,995} & {\bf 2,443} & {\bf 21,006} & {\bf 759.8\%}\\
\hline\hline
Old PDF & 22,081 & 6,306 & 6,475 & 2.7\%\\
\hline%\hline
New PDF & 1,428 & 32 & 170 & 431.2\%\\
\hline
{\bf PDF Total} & {\bf 23,509} & {\bf 6,338} & {\bf 6,645} & {\bf 4.8\%}\\
\hline%\hline
\end{tabular}
%}
\caption{Effectiveness Results.}
\label{table:effectiveness}
\vspace{-0.2in}
\end{table}
}

\begin{table*}[!t]
\centering
\begin{tabular}{|l|l|l|l|l|l|l|}
\hline
{\bf Sample Set} & {\bf Total}   & \specialcell{\bf without \\ \codename} & \specialcell{\bf with \\ \codename} & {\bf Improvement} & \specialcell{\bf Detected\\ \bf By Both} & \specialcell{ \bf Missed\\ \bf With \codename} \\ \hline
Old HTML           & 66,325          & 193                                                   & 357                                                & 84.9\%              & 193            & 0                 \\ \hline
New HTML           & 106,018         & 2,250                                                 & 20,649                                             & 817.3\%             & 2250           & 0                 \\ \hline
{\bf HTML Total} & {\bf 172,995} & {\bf 2,443}                                         & {\bf 21,006}                                     & {\bf 759.8\%}     & {\bf 2443}           & {\bf 0}                 \\ \hline
Old PDF            & 22,081          & 6,306                                                 & 6,475                                              & 2.7\%               & 6306           & 0                 \\ \hline
New PDF            & 1,428           & 32                                                    & 170                                                & 431.2\%             & 32             & 0                 \\ \hline
{\bf PDF Total}  & {\bf 23,509}  & {\bf 6,338}                                         & {\bf 6,645}                                      & {\bf 4.8\%}       & {\bf 6338}           & {\bf 0}                 \\ \hline
\end{tabular}
\caption{Effectiveness Results.}
\label{table:effectiveness}

\end{table*}

\subsection{Effectiveness}\label{sec:effectiveness}
\setcounter{paragraph}{0}

\hu{add data for \codename without object retyping.}

For the evaluation of effectiveness, we would like to demonstrate that \codename can indeed help the malicious JavaScript code analysis by performing efficient forced execution. In order to achieve that, we utilize our malicious HTML and PDF sample sets and run the sample sets against jsunpack both with or without \codename for the evaluation. In the interest of showing how useful our faked object retyping is, we also conduct another experiment that disables the retyping and only keeps the reference error recovery component and path exploration component.

\paragraph{Experimental Results}

Table~\ref{table:effectiveness} illustrates the experimental results for effectiveness. It demonstrates that \codename could greatly improve the detection rate for JavaScript analysis. We can see detection rate improvements of 759.84\% and 4.84\% for HTML and PDF samples, respectively, when using \codename-extended jsunpack instead of the original version for analysis. And all the samples detected by original jsunpack are also flagged by \codename-extended jsunpack.  We further break down the numbers into old and new sample sets and perceive that the extended version could perform much better than original jsunpack in analyzing new samples. For new HTML samples, jsunpack with \codename is able to detect 817.3\% more samples while for old samples, the number is 84.97\%. Similar results are also observed for PDF samples. After manual inspection, we confirmed that this is because many of the old samples have been analyzed for quite sometime and jsunpack already has the signatures stored in its database, leaving only a small margin for \codename to improve upon. For the faked object retyping evaluation, we reran the test using 106,018 new HTML malicious samples with retyping component disabled. The result shows that only 8,677 samples can be detected by \codename in contrast to 20,649 with retyping enabled. This result reveals the usefulness of our faked object retyping component during analysis. Nevertheless, through our experiments, we are able to draw the conclusion that \codename is quite effective for boosting the effectiveness of JavaScript analysis.

\hu{explain the confusion from reviewerD about the 96\% number and 98.33\% in section 6. }
\paragraph{Number of Paths Explored}

\begin{figure}[t]
  \centering
  \includegraphics[width=\linewidth]{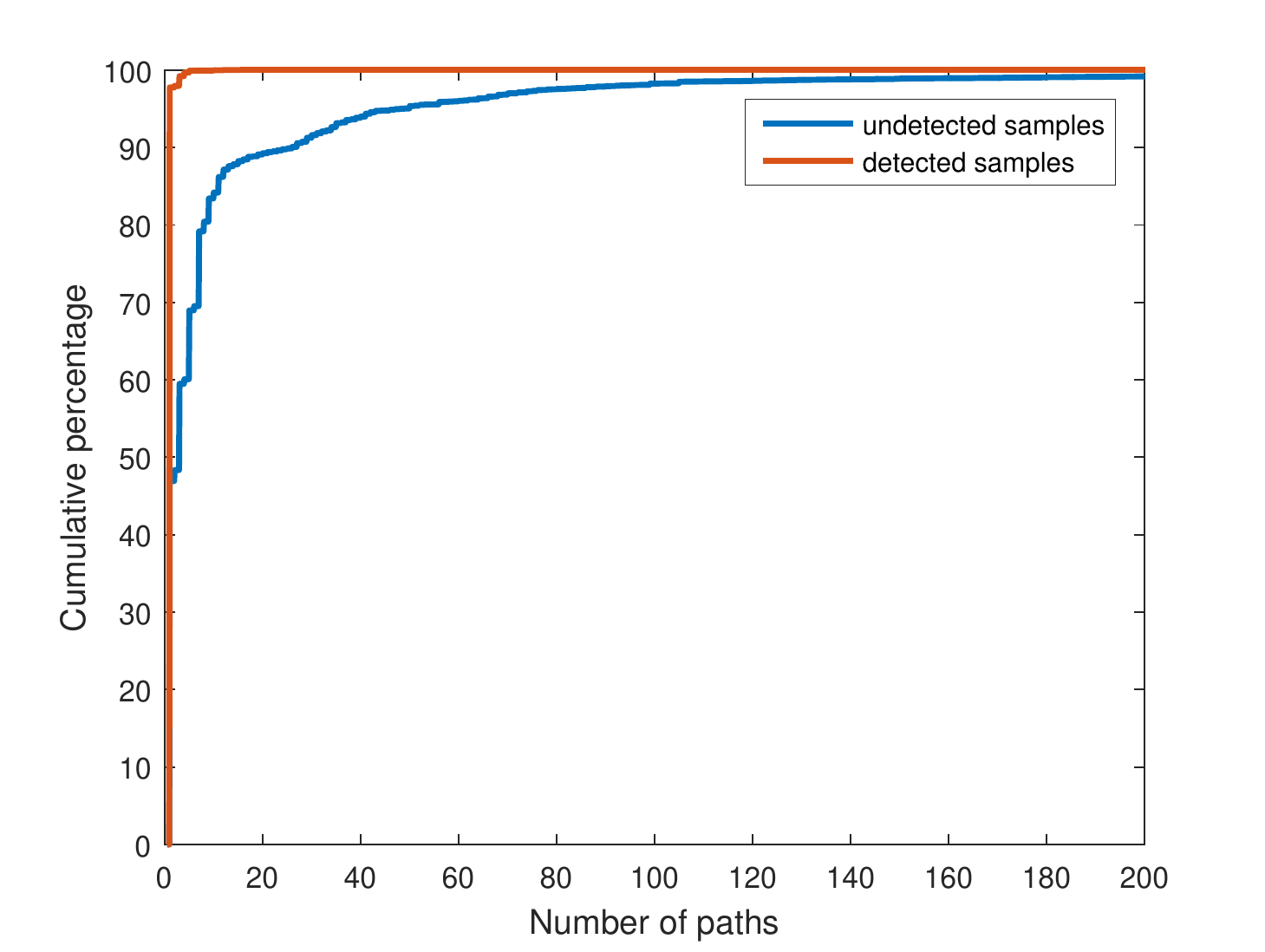}\\
  \caption{Num of Path Exploration during Analysis.}\label{fig:pathnum}
\end{figure}

Potentially, there may be a large number of paths that exist inside of a single JavaScript program. The effectiveness and efficiency of \codename are closely related to the number of paths explored during analysis. Hence, we would like to show some statistics on the number of paths that \codename explored during analysis. %Since the number of paths explored for undetected samples makes no sense as it merely reflects the total number of paths in the tested JavaScript program, we thereby only show the results for detected HTML and PDF samples. 

The result depicted in Figure~\ref{fig:pathnum} shows that \codename is able to detect the maliciousness of samples with a limited number of path explorations. An interesting observation is that over 96\% of the samples were detected by exploring only a single path.  Even though most of the analysis for detected samples can be finished by exploring just one path, the path exploration of \codename is still essential. Note that 98\% of the samples missed by the default jsunpack, but detected by the \codename-extended version, explore at least two paths. So, the analysis could still receive an enormous benefit from \codename in terms of path exploration.\yue{explain a little more why. Would a fail-free engine get almost the same results} Please refer to the Section 6 Case Study for more details on this topic. As for any undetected samples, \codename will explore the entire code space during analysis, which requires a larger amount of path exploration and longer analysis runtime.

\subsection{Runtime Performance}
\setcounter{paragraph}{0}

In this section, we evaluate the runtime performance of \codename by using our malicious and benign datasets with a comparison between the original jsunpack and the \codename-extended version.

\ignore
{
\begin{table*}[t]
%\vspace{-0.1in}
\small
\centering
%\resizebox{\linewidth}{!}{
\begin{tabular}{|l|l|l|l|l|}
\hline
{\bf } & \multicolumn{2}{c|}{\bf {w/o \codename}} & \multicolumn{2}{c|}{\bf {with \codename}}\\
\hline
{\bf Category} & {\bf avg. Runtime} & {\bf $80^{th}$ Percentile} & {\bf avg. Runtime} & {\bf $80^{th}$ Percentile}\\
\hline\hline
Detected HTML & 2.69s & 2.76s & 0.98s & 0.91s\\
\hline%\hline
Detected PDF & 7.24s & 9.95s & 8.87s & 6.98s\\
\hline\hline
Undetected HTML & 1.13s & 1.41s & 16.08s & 9.61s\\
\hline%\hline
Undetected PDF & 1.37s & 1.49s & 7.97s & 7.40s\\
\hline
\end{tabular}
%}
\caption{Runtime Overhead.}
\label{table:overhead}
\vspace{-0.2in}
\end{table*}
}

\paragraph{Runtime for Detected Samples}

\begin{figure*}[t]
\minipage{0.5\textwidth}
  \centering
  \includegraphics[width=1\linewidth]{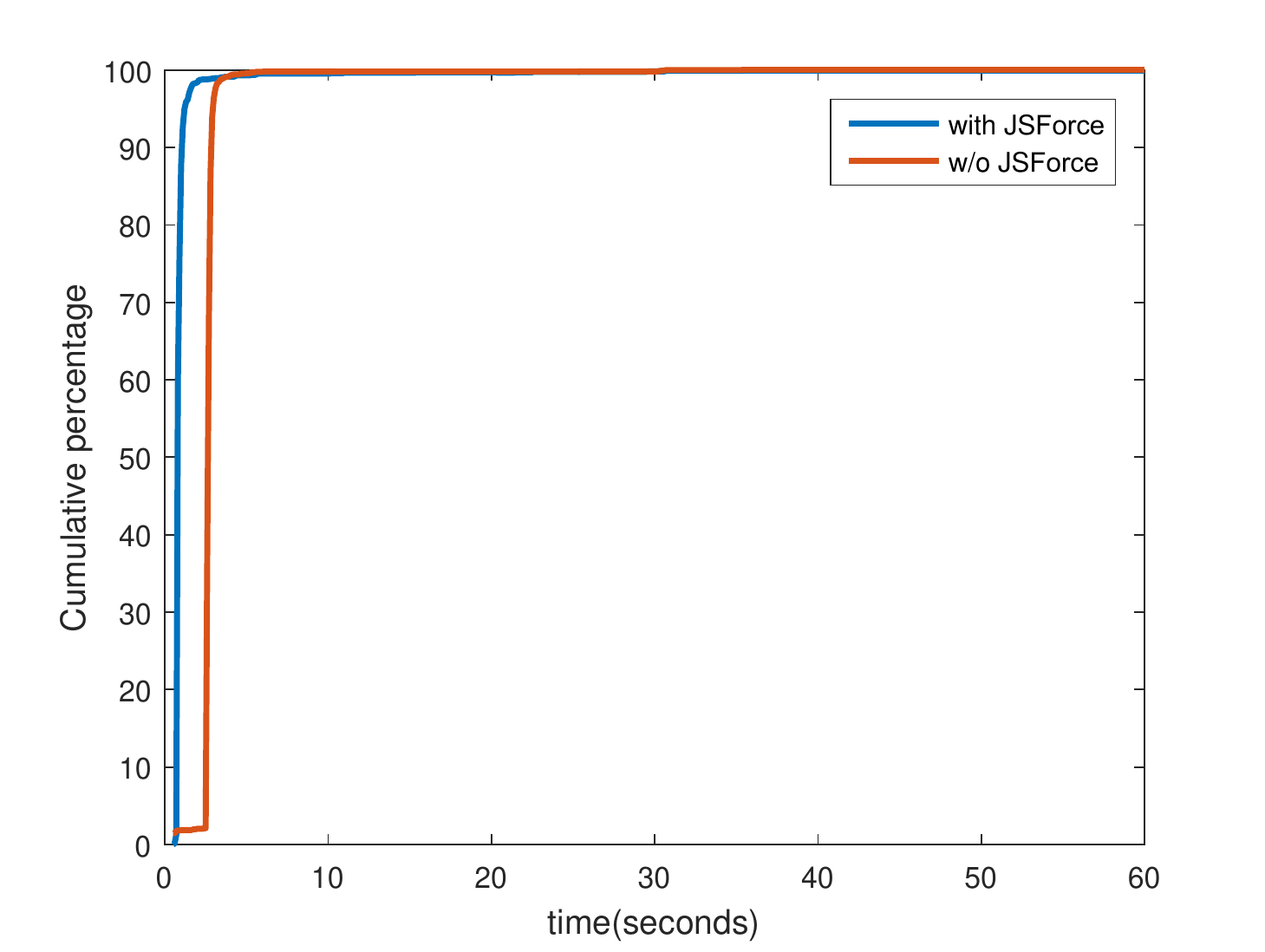}\\
  \caption{Runtime for Detected HTML samples.}\label{fig:detectedhtmlruntime}
\endminipage\hfill
\minipage{0.5\textwidth}
  \centering
  \includegraphics[width=1\linewidth]{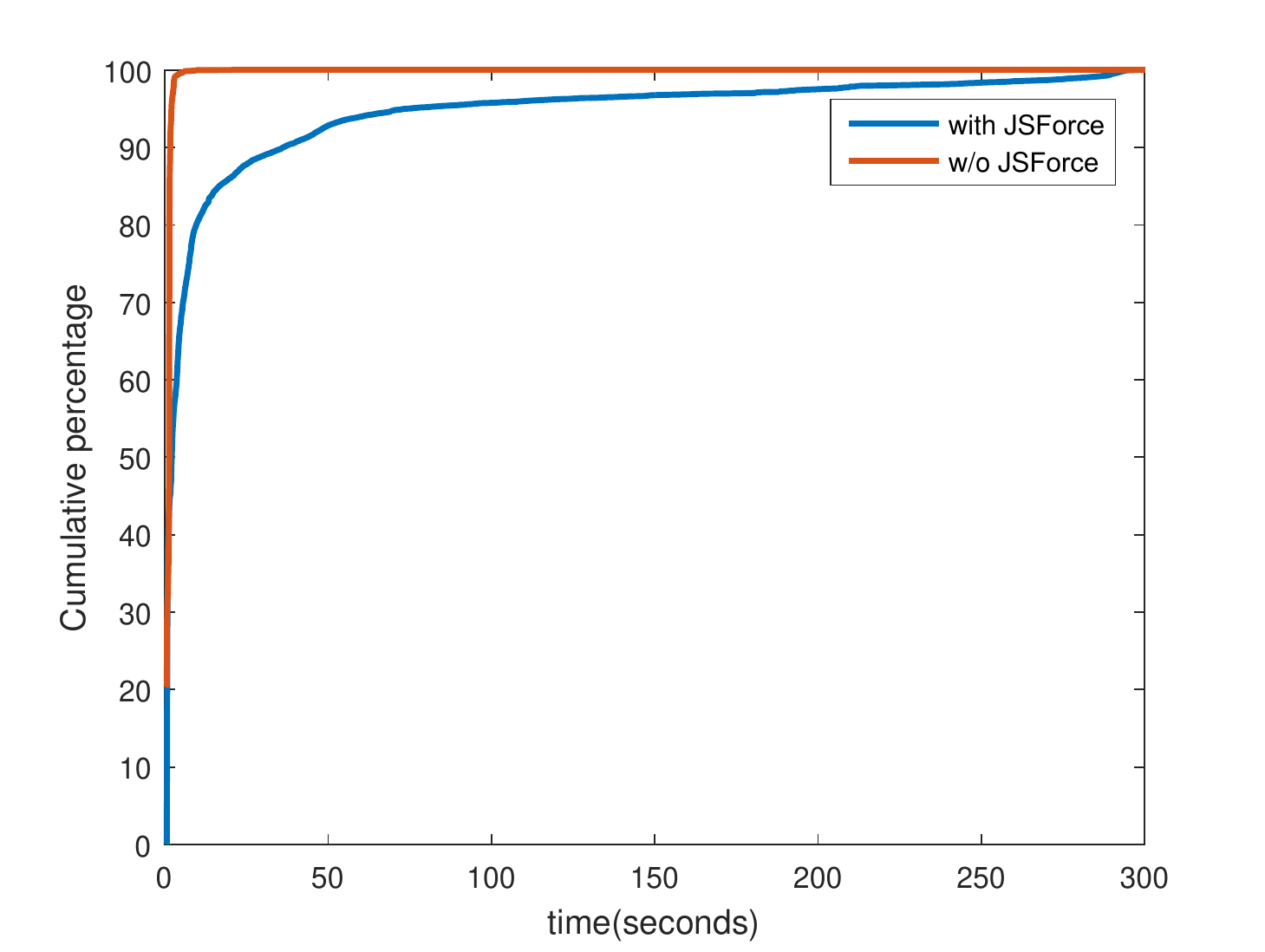}\\
  \caption{Runtime for Undetected HTML samples.}\label{fig:undetectedhtmlruntime}
  \endminipage\hfill
\end{figure*}

\begin{figure*}[!htb]
\minipage{0.5\textwidth}
  \centering
  \includegraphics[width=1\linewidth]{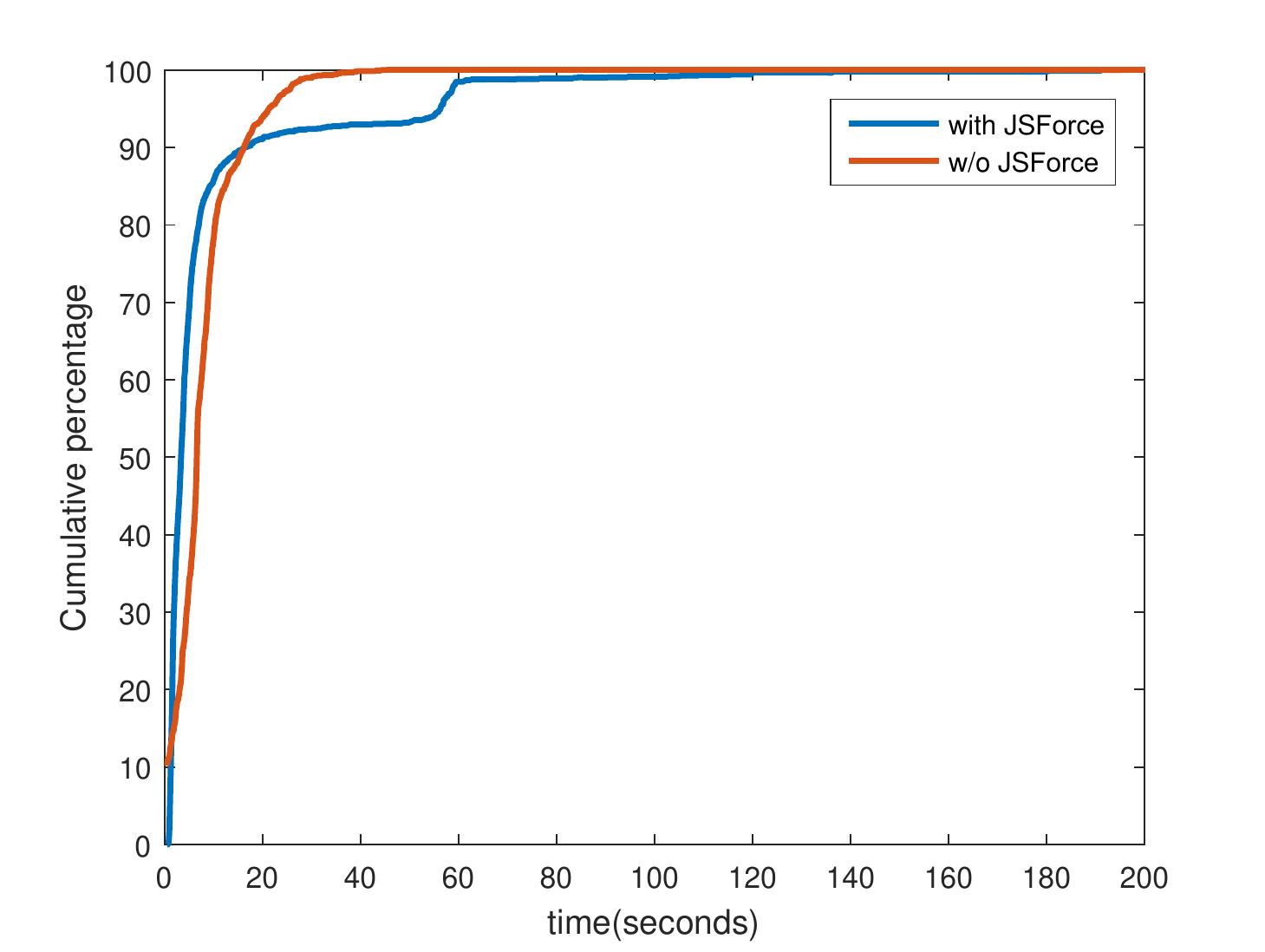}\\
  \caption{Runtime for Detected PDF samples.}\label{fig:detectedpdfruntime}
\endminipage\hfill
  \minipage{0.5\textwidth}
  \centering
  \includegraphics[width=1\linewidth]{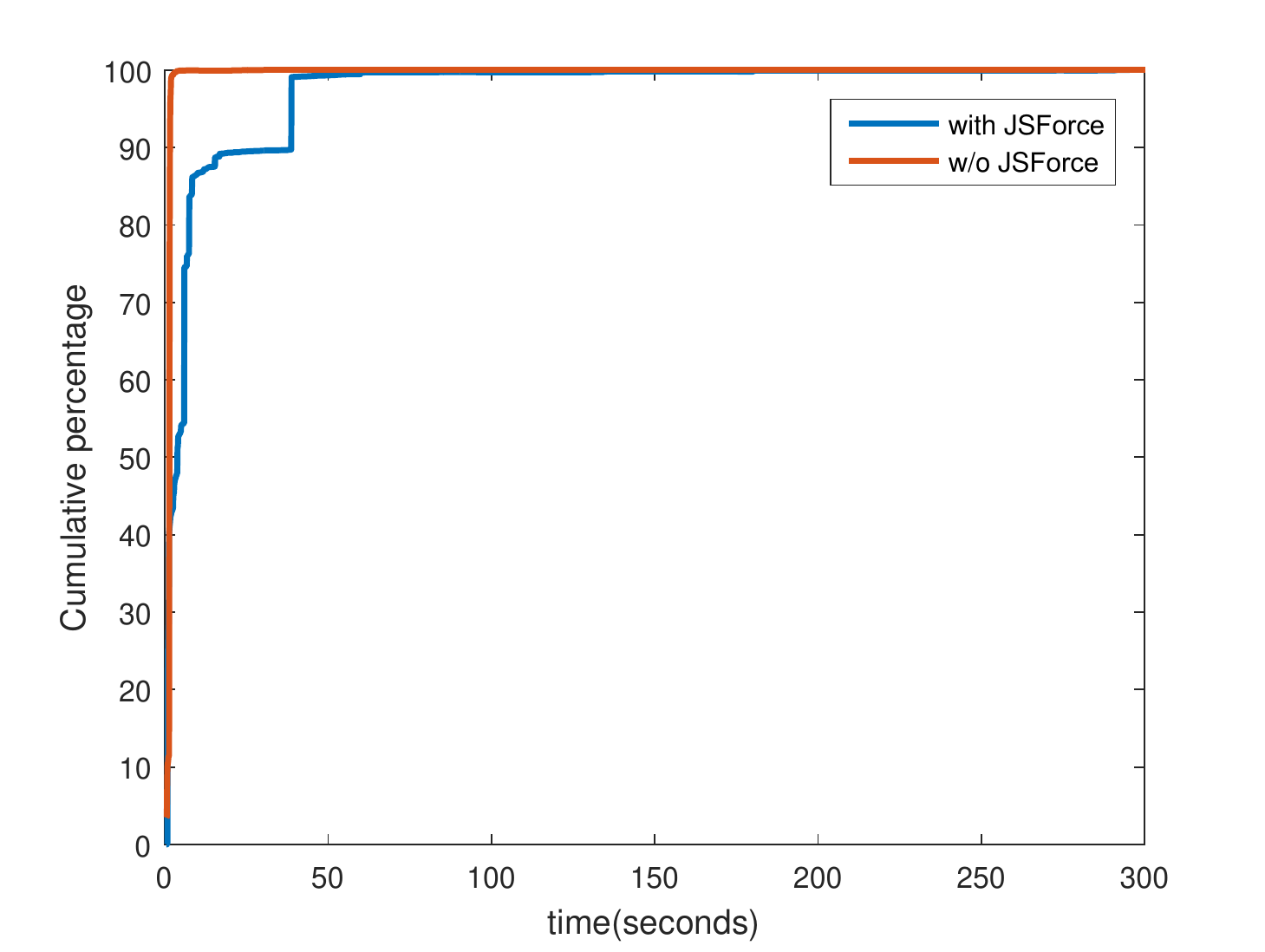}\\
  \caption{Runtime for Undetected PDF samples.}\label{fig:undetectedpdfruntime}
  \endminipage\hfill
\end{figure*}

In this section, we compare the runtime performance using the HTML and PDF samples that can be  detected by jsunpack both with and without \codename. The reason why we chose this sample set is that we wished to observe whether the \codename-extended version can achieve efficiency comparable to the original jsunpack when using a detectable malicious sample. The results are displayed in Figures~\ref{fig:detectedhtmlruntime} and~\ref{fig:detectedpdfruntime}. The results conclude that \codename-extended version has better runtime performance than jsunpack for over 90.9\% of HTML and 83.6\% of PDF samples. This conclusion is quite surprising as the \codename-extended version tends to explore multiple paths while jsunpack only probes for one.

In theory, jsunpack should have better runtime performance. However, after investigation, we found that many of the JavaScript samples require specific system configurations (such as specific browser kernel version) to run. As a result, when jsunpack performs analysis, it will run the JavaScript programs under multiple settings. This results in multiple executions, which take additional time to complete. In contrast, the \codename-extended version handled this issue with forced execution, resulting in better runtime performance in practice.

\paragraph{Runtime for Undetected Samples}

Figures~\ref{fig:undetectedhtmlruntime} and~\ref{fig:undetectedpdfruntime} show the runtime performance of \codename for undetected samples. We empirically set the time limit to be 300 seconds in consequence of the fact that experiment shows almost all (99.6\%) HTML and PDF samples can be analyzed within 300 seconds. As demonstrated in the figures, the average analysis runtime for HTML and PDF samples are 12.02 and 8.15 seconds, while the analysis for a majority (80\%) of HTML samples and PDF samples are finished within 8.54 and 7.4 seconds, respectively. When compared with the original jsunpack, the \codename-extended version achieves an average runtime of 16.08 seconds and 7.97 seconds for undetected HTML and PDF samples while jsunpack finishes execution in 1.13 seconds and 1.37 seconds, correspondingly. Our conclusion from these experiments are that the performance overhead of \codename is quite reasonable and can certainly meet the requirements of large scale JavaScript analysis.

\ignore{
\begin{table*}[]
\centering
\begin{tabular}{|l|l|l|l|l|}
\hline
               & Conf. w/o Rozzle & Conf. With Rozzle & DetectedByBothConf & {MissedByRozzle-extendedConf.} \\ \hline
offline-Nozzle & 1662       & 11559       & 1178             & 484(29\%)                            \\ \hline
online-Nozzle  & 74         & 224         & 50               & 24(32\%)                             \\ \hline
online-Zozzle  & 2735       & 2660        & 2510             & 225(8\%)                            \\ \hline
\end{tabular}
\caption{Detection Results With/Without Rozzle-extended Configuration}
\label{table:rozzleResults}
\end{table*}
}

% Please add the following required packages to your document preamble:
% \usepackage{multirow}
\begin{table*}[]
\centering
\begin{tabular}{|l|l|l|l|l|l|}
\hline
AnalysisSystem          & SampleSet & Conf. w/o Rozzle & Conf. With Rozzle & DetectedByBothConf & MissedByRozzle-extendedConf. \\ \hline
\multirow{2}{*}{Nozzle} & Offline   & 1,662             & 11,559             & 1,178               & 484(29\%)                        \\ \cline{2-6} 
                        & Online    & 74               & 224               & 50                 & 24(32\%)                         \\ \hline
Zozzle                  & Online    & 2,735             & 2,660              & 2,510               & 225(8\%)                         \\ \hline
\end{tabular}
\caption{Detection Results With/Without Rozzle-extended Configuration}
\label{table:rozzleResults}
\end{table*}

\subsection{\codename vs. Rozzle}
\setcounter{paragraph}{0}

Ideally, we would like to perform a head-to-head comparison between \codename and Rozzle using the same dataset. Unfortunately, it is impossible given that neither the Rozzle system nor the dataset used by Rozzle is available for evaluation. It is also nontrivial to implement Rozzle by ourselves. Nevertheless, we can still highlight several advantages of \codename over Rozzle, from the experimental results reported in that paper.

%Both \codename and Rozzle can be used as the amplifier technique to increase the code coverage of JavaScript analysis systems. However, \codename advantages Rozzle in several aspects. 

First, while Rozzle-extended analysis system does, \codename-extended analysis system does not miss samples detected by the original analysis system. Table~\ref{table:rozzleResults} summarizes the detection results presented in Rozzle paper. Using Rozzle, the experiments extend two malicious JavaScript detection systems - Nozzle~\cite{nozzle:usenix09} and Zozzle~\cite{curtsinger2011zozzle}, and then compare the detection results with the original system using one offline sample set and one online sample set. For the offline experiment, with Rozzle, Nozzle can detect 11,559 samples and gains a significant improvement(11,559 vs. 1,662) over original Nozzle. But it misses 484 (29\%) samples which can be detected by original Nozzle. For online experiments, Rozzle-extended configuration also misses 24 (32\%) for Nozzle, 225 (8\%) for Zozzle respectively. Rozzle paper argues this is because that the runtime errors, introduced when infeasible paths are executed, terminates the execution before the malicious behaviors are exposed. However, since \codename only collects the path information and no changes are made on the path when the sample is first executed, no runtime errors are introduced by \codename. Thus as demonstrated in Section~\ref{sec:effectiveness}, \codename-extended analysis system can detect all the samples identified by original analysis system while providing the same magnitude improvement as Rozzle's.

Second, \codename can still function even when the environment setup is incomplete, thanks to the forced execution model (Section~\ref{sec:forcedexecution}), whereas Rozzle may fail due to the runtime errors. This is especially important for low-interaction honey clients like jsunpack. Those low-interaction honey clients emulate the behaviors of browsers or PDF readers, and it is quite challenging to construct a complete environment setup for the tested samples. As discussed in Section~\ref{sec:casestudyb}, of the malicious samples missed by jsunpack, 96.5\%  are  because of the runtime errors caused by incomplete emulation of the running environments for JavaScript code. Since low-interaction honey clients are widely deployed in industry, we argue that \codename would benefit the industry more than Rozzle.  

Third, as discussed in the limitation part of Rozzle paper, Rozzle is less effective for the case that the evasive code triggers the malware execution only when a user interaction occurs, or when a timer fires. We searched the samples missed by jsunpack with the keywords like ``onclick'' or ``settimeout''. we found that  80.6\% of them deploy timers or user interaction callbacks. \codename's path exploration algorithm discovers the callback functions during the execution, and invokes them after the current run terminates. However, Rozzle may miss the malicious code hidden in callback functions.  

Fourth, Rozzle cannot handle latest fingerprinting techniques discussed in Section~\ref{sec:challenge}. While we have not found samples deploying these techniques in our dataset, we believe that the attacker will deploy those new fingerprinting techniques with the advancement of anti-evasion techniques in the future. So \codename is one step ahead of the attacker.

\section{Case Study}

\begin{figure}[h]
\centering

\noindent\begin{minipage}{.22\textwidth}
\begin{lstlisting}[numbers=none, frame=single,basicstyle=\fontsize{7}{13}\selectfont]{[a]}
try{++(document["body"]);
} catch(e) {[shellcode]}
\end{lstlisting}
\end{minipage}\hfill
\noindent\begin{minipage}{.22\textwidth}
\begin{lstlisting}[numbers=none, frame=single,basicstyle=\fontsize{7}{13}\selectfont]{[b]}
window.addEvent('load', 
  function() {[shellcode]})
\end{lstlisting}
\end{minipage}\hfill%

\noindent\begin{minipage}{.48\textwidth}
\begin{lstlisting}[numbers=none, frame=single,basicstyle=\fontsize{7}{13}\selectfont]{[c]}
function frmAdd() {
  var ifrm = document.createElement('iframe');
  ifrm.style.position='absolute';
  document.body.appendChild(ifrm);};
frmAdd();
\end{lstlisting}
\end{minipage}\hfill

\noindent\begin{minipage}{.48\textwidth}
\begin{lstlisting}[numbers=none, frame=single,basicstyle=\fontsize{7}{13}\selectfont]{[d]}
var se_show_newupdates = new Hash.Cookie('se_show_newupdates', {duration: 3600});
\end{lstlisting}
\end{minipage}\hfill

\noindent\begin{minipage}{.48\textwidth}
\begin{lstlisting}[numbers=none, frame=single,basicstyle=\fontsize{7}{13}\selectfont]{[e]}
gapi.load("gapi.iframes", function() {[...]});
\end{lstlisting}
\end{minipage}\hfill

\caption{Case Study Samples}\label{fig:casestudy}
\end{figure}

To better understand the benefits of \codename, we conducted a case study on 10,975 unique JavaScript code samples missed by jsunpack but detected by \codename-extended version. The reasons of failed detection using jsunpack can be divided into the following two categories. 

\paragraph{Malicious code branch is not triggered}

Of those 10,975 samples missed by jsunpack, we found that 10,792 (98.33\%) samples are explored by at least two paths when using \codename. Although jsunpack attempts to run the sample several times with different configurations to increase the chance of triggering the malicious code branch, it is usually ineffective to do so. This is because it is impossible to emulate every single combination of browser/PDFReader/plugin. In contrast, \codename can explore these paths regardless of the configuration.

The sample in Figure~\ref{fig:casestudy}(a) hides malicious code within a {\tt catch} block. The attacker attempts to increase the value {\tt document[body]} as a number, which will raise an exception when executed within a real browser. However, it does not raise an exception in jsunpack since its SpiderMonkey engine returns {\tt NaN} for this operation. In fact, the V8 engine used in \codename also exhibits the same behavior as SpiderMonkey. But, the {\tt catch} block is triggered by the path exploration process, so the malicious behavior is still revealed. 

Other samples hide code within event callbacks. For instance, the sample in Figure~\ref{fig:casestudy}(b) registers a callback function using the {\tt window.addEvent} function. jsunpack fails to invoke the callback function due to the incorrect definition of {\tt window.addEvent} used by jsunpack. At runtime, \codename identifies {\tt window.addEvent} as a callback registration function because an anonymous function is passed to it as the parameter. Then, this anonymous function is queued and invoked at the end of execution.

\hu{Better way to present the runtime errors. how many runs jsunpack execute for every sample, for every run, what kind of errors are introduced }

\paragraph{Execution fails due to runtime errors}\label{sec:casestudyb}

Another reason why jsunpack may fail to detect malicious JavaScript code is that the execution can fail due to runtime errors. As we conducted the evaluation, \ignore{about half of JavaScript code executions (31630 out of 65850\footnote{Each sample was executed six times by jsunpack due to different configuration settings}) failed due to three types of runtime errors: {\small\tt TypeError}, {\small\tt ReferenceError} and {\small\tt SyntaxError}. More specifically, we encountered 10,730 {\small\tt TypeError}, 10,510 {\small\tt ReferenceError} and 10,390 {\small\tt SyntaxError} exceptions respectively.}only 230 out of the 10,975 samples could be executed without any runtime errors under the six configurations. Moreover, 10,592 out of 10,975 (96.5\%) failed all six configurations, rendering jsunpack completely useless when facing them. These exceptions terminate the execution before the malicious code is executed. The raised exceptions are because of the inaccurate emulation of the running environment for JavaScript code. Examining these exceptions can help security researchers improve jsunpack by supplying more precise emulation environment, which is another benefit that \codename can provide.

%When analyzed with jsunpack, 10,730 (97.77\%) samples raise {\small\tt TypeError} exceptions, 10,510(95.76\%) raise {\small\tt ReferenceError} exceptions, and 10390(94.67\%) raise {\small\tt SyntaxError} exceptions.  These exceptions terminate the execution before the malicious code is executed. The raised exceptions are because of the inaccurate emulation of the running environment for JavaScript code. Examining these exceptions can help security researchers improve jsunpack by supplying more precise emulation environment, which is another benefit that \codename can provide.

One interesting thing about jsunpack is that it tries to fix {\small\tt ReferrenceError} by providing a definition for this undefined object once {\small\tt ReferrenceError} is captured.  While this fix eliminates the {\small\tt ReferrenceError}, it often introduces {\small\tt SyntaxError} or {\small\tt TypeError} at runtime. {\tt ifrm.style} is not defined in the sample in Figure~\ref{fig:casestudy}(c).  So jsunpack generates code {\tt var\ ifrm.style = 1} for this sample. Unfortunately, it contains an unexpected token dot. This raises a {\small \tt SyntaxError} exception. Another way to improve this is to assign {\tt ifrm.style} an {\tt Object} so that {\small \tt SyntaxError} is avoided and {\tt ifrm.style} can be typed following the typing rules of the JavaScript engine. However, as discussed in Section~\ref{sec:forcedexecution}, this can still cause an exception or lead to unnecessary loss of precision. This case demonstrates the advantage of type inference model deployed by \codename. Although \codename cannot tolerate {\small\tt SyntaxError}, the type inference model guarantees no further {\small\tt TypeError} or {\small\tt SyntaxError} will be introduced.

The sample in Figure~\ref{fig:casestudy}(d) raises a {\small\tt TypeError} exception since {\tt Hash.Cookie} is not a constructor. Another sample in Figure~\ref{fig:casestudy}(e) also raises a {\small\tt TypeError} exception because {\tt gapi.load} is not a function. \codename can avoid this by applying faked object retyping technique. From another perspective, these two cases manifest the weakness of jsunpack that  {\tt Hash.Cookie} and {\tt gapi.load} are not correctly defined. Therefore, as another application, \codename can be used to evaluate the weakness of dynamic JavaScript analysis systems, so security researchers can further improve the systems respectively.

\section{Limitations}
\setcounter{paragraph}{0}

If the syntax of the tested JavaScript code is not correct, \codename drops the analysis immediately. The forced execution can introduce syntax error under some cases. For instance, the parameter of {\tt eval} is supposed to be correct JavaScript code. When the parameter is calculated from faked strings created by \codename, the parameter may become syntax incorrect for {\tt eval}. In the future, we expect to develop techniques~\cite{barnard1982hierarchic} to automatically fix the syntax error to enable maximized execution of the code.
\begin{figure}
\centering

\begin{lstlisting}[multicols=2]
var f = function() {
    if (true) {
        function g(){return 1;}
    } else {
        function g(){return 2;}
      };
    function g(){return 3;};
    return g();
    function g(){return 4;}
}
\end{lstlisting}
\caption{A JavaScript Sample Interpreted Differently by Different JavaScript Engines}
\label{fig:implDif}
\end{figure}

While JavaScript language has the official specification from the ECMAScript community~\cite{ecmascript}, the implementation of the language on different JavaScript engines differs slightly because of the complex features and rapid evolving of JavaScript language. The attacker can exploit this weakness to create a deliberate script which exhibits differently on \codename to evade the analysis. Maffeis et al.~\cite{maffeis2008operational} discussed such an example presented in Figure~\ref{fig:implDif}. This code defines a function {\tt f} whose behavior is given by one of the declarations of {\tt g} inside the body of the anonymous function that returns {\tt g}. However, different implementations disagree on which declaration determines the behavior of {\tt f}. Specifically, a call to {\tt f()} should return 4, according to ECMA specification. SpiderMonkey returns 4, while Rhino and Safari return 1, and JScript and the ECMA4 reference implementations return 2. Attackers can leverage these differences to hide the decoding key and evade analysis. To counter this, we can implement \codename on top of different JavaScript engines, such as SpiderMonkey~\cite{spidermonkey} and Chakra~\cite{chakra}.

The current path exploration algorithm can efficiently explore most of the sample in a decent time. However, there are still some cases that take a considerable length of time to finish. To exploit this limitation, attackers may place the malicious code deep in the code logic, such that \codename could not reach it within a predefined duration. Note that this limitation is not unique for \codename. All the path exploration techniques share the same limitation. We leave it as future work to develop better path exploration algorithms and search heuristics.
%In the future, we expect to develop a more efficient path exploration algorithm by introducing tainting to guide exploration towards only the execution paths related to input.

\ignore{

\codename can be evaded by techniques that do not need control-flow branches, e.g., those based on browser or JavaScript quirks. For example, the property $window.innerWidth$ is defined in FireFox and Chrome but undefined in Internet Explorer. Therefore, a malicious code that computed a decoding key from $window.innerWidth$  would obtain a different result in Firefox/Chrome and IE, and could be used to decode malicious code in specific browsers. \codename will not trigger the malicious code path in such cases and can be evaded. 

%However, we can model those differences in a predefined JavaScript file to handle such cases for \codename.

\begin{figure}
\centering

\begin{lstlisting}[frame=single]
xxxxx = 'ev';
yyyyy = 'al';
zzzzz = xxxxx + yyyyy;
aaaaa = app;
try {} catch (e) {
    zzzzz = 1;
    aaaaa = 1;
}
try {
    d = nothis_nothis;
    zzzzz = 1;
    aaaaa = 1;
} catch (e) {}
aaaaa[zzzzz]('ddddd' + 'dd=une' + 'sca' /**/ + /**/ 'pe;');
\end{lstlisting}

\caption{The Case of Evading \codename}
\label{fig:evasioncase}
\end{figure}

Figure~\ref{fig:evasioncase} presents an example that can bypass \codename. Using normal JavaScript engine,  lines 6-7 and lines 11-12 will not be executed so that $aaaaa$ and $zzzzz$ can be correctly initialized. However, with \codename, line 10 will not raise the exception since $d$ will be initialized as $FakedObject$. The exponential path exploration can handle this case by exploring all the possible path combinations. But it is not feasible in practice. Our solution is that we  execute the sample without forced execution when we collect the path predicates at the beginning.

}

\section{Related Work}\label{sec:related}
\setcounter{paragraph}{0}

\paragraph{Malicious JavaScript Analysis} 
In the last few years, there have been a number of approaches to analyzing JavaScript code. They can be roughly divided into two categories-static approach, dynamic approach.

{\em Static Approach.}  Several systems have focused on statically analyzing JavaScript code to identify malicious web pages~\cite{feinstein2007caffeine, likarish2009obfuscated,seifert2008identification, curtsinger2011zozzle}. ZOZZLE~\cite{curtsinger2011zozzle}, in particular, leverages features associated with AST context information (such as, the presence of a variable named shellcode in the context of a loop), for its classification. Since dynamic features of JavaScript plague the static analysis , researchers try to model those features to improve the static analysis result~\cite{politz2015adsafety, taly2011automated, bandhakavi2010vex}.

{\em Dynamic Approach.} Dynamic analysis is widely deployed to expose behaviors of obfuscated JavaScript code. Previous work~\cite{cova10:wepawet,Lu:2012,hartstein2009jsunpack} execute JavaScript using an emulated JavaScript running environment and acquire de-obfuscated JavaScript code. To de-obfuscate malicious JavaScript code,  Gen et al. ~\cite{Lu:2012} simplify the obfuscated JavaScript code by preserving the semantics of the observational equivalence. JSGuard~\cite{jsguard:gu} proposed a methodology to detect JavaScript shellcode that fully uses JavaScript code execution environment information with low false negative and false positive.  Liu et al.  ~\cite{liu2014detecting} propose a context-aware approach for detection and confinement of malicious JavaScript in PDF by inserting context monitoring code into a document. To analyze JavaScript code with cloaking , Kolbitsch et al.~\cite{kolbitsch2012rozzle} uncover environment-specific malware by exploring multiple execution paths within a single execution. \codename can benefit the dynamic analysis in terms of improved code coverage and tolerance of invalid host environment model.

Researchers also try to combine static and dynamic code features to identify malicious JavaScript programs(Cujo~\cite{rieck2010cujo}). More precisely, Cujo processes the static program and traces of its execution into q-grams that are classified using machine learning techniques. Symbolic execution~\cite{saxena2010symbolic} is also explored for malicious JavaScript analysis. 
%Revolver~\cite{kapravelos2013revolver} are proposed to detect evasive web-based malware. It leverages the observation that two scripts that are similar should be classified in the sameway by web malware detectors (either both scripts are malicious or both scripts are benign); differences in the classification may indicate that one of the two scripts contains code designed to evade a detector tool. However , it can not detect 0 day attack since it relies on known sample set to identify the difference.

%Rozzle~\cite{kolbitsch2012rozzle} have studied the “fragility” of malicious code, i.e., its dependence for correct execution on the presence of a particular execution environment (e.g., specific browser and plugins versions).  They propose ROZZLE, a system that explores multiple execution paths in a program, thus bypassing environment checks. Rozzle only detects fingerprinting that leverages control flowbranches and depends upon the environment. It can be evaded by techniques that do not need control-flow branches.

%JSUnpack~\cite{hartstein2009jsunpack} emulates browser functionality when visiting a URL. It is designed to detect exploits that target browser and browser plug-in vulnerabilities. 

\paragraph{Forced Execution} Researchers have proposed to force branch outcomes for different security applications. X-Force~\cite{xforce} can force the binary to execute and explore different execution paths requiring no inputs or proper environment. iRis~\cite{deng2015iris} employs forced execution technique to expose the private API abuses in iOS applications. Forced execution was also proposed to identify kernel-level rootkits]~\cite{wilhelm2007forced}, expose hidden behavior in Android apps~\cite{johnson2013forced, wang2012exposing}. To the best of our knowledge, \codename is the first work to enable forced execution on JavaScript for malware detection.
\section{Conclusion}

In this paper, we presented the design and implementation of a novel JavaScript forced execution engine named \codename which enables non crashable execution model  while ensuring complete code coverage. We evaluated \codename using a large number of HTML and PDF samples. Experimental results showed that by adopting our forced execution engine, the malicious JavaScript detection rate can be greatly improved without any noticeable false positive increase and the runtime overhead was generally neglectable. \codename is made publicly available at ~\cite{servicelink} as an online service and will release the source code to the security community upon the acceptance for publication.

{\footnotesize
\bibliographystyle{acm}
\bibliography{all}
}

%\appendix
%\input{appendix}

\end{document}